\documentclass[apj]{emulateapj}

\def\spose#1{\hbox to 0pt{#1\hss}}
\def\simlt{\mathrel{\spose{\lower 3pt\hbox{$\mathchar"218$}}
     \raise 2.0pt\hbox{$\mathchar"13C$}}}
\def\simgt{\mathrel{\spose{\lower 3pt\hbox{$\mathchar"218$}}
     \raise 2.0pt\hbox{$\mathchar"13E$}}}
\slugcomment{Accepted for publication in ApJ}
\usepackage{epsf}
\usepackage{natbib}
\usepackage{url}
\font\smcap=cmcsc10

\newcommand{\kms}{\,km~s$^{-1}$}
\newcommand{\degree}{$^\circ$}
\newcommand{\nai}{Na\,{\smcap i}}
\newcommand{\caii}{Ca\,{\smcap ii}}

\newcommand{\vio}{$(V-I)_0$}
\newcommand{\ivi}{($I,\,V-I$)}

\newcommand{\feh}{$\rm[Fe/H]$}
\newcommand{\fehp}{$\rm[Fe/H]_{phot}$}

\newcommand{\olkhd}{$\langle L_i\rangle$}

\newcommand{\mvsph}{$\langle v\rangle^{\rm sph}$}
\newcommand{\sigvsph}{$\sigma^{\rm sph}_v$}
\newcommand{\mvsb}{$\langle v\rangle^{\rm sub}$}
\newcommand{\sigvsb}{$\sigma^{\rm sub}_v$}
\newcommand{\rproj}{$R_{\rm proj}$}

\shorttitle{Surface Brightness Profile of M31's Halo}
\shortauthors{Gilbert et~al.}

\begin{document}
\bibliographystyle{apj}

\title{Global Properties of M31's Stellar Halo from the SPLASH Survey.  I. Surface Brightness Profile\footnote{The data presented herein were obtained at the W.M. Keck Observatory,
which is operated as a scientific partnership among the California
Institute of Technology, the University of California and the National
Aeronautics and Space Administration. The Observatory was made
possible by the generous financial support of the W.M. Keck
Foundation.}}

\author{
Karoline~M.~Gilbert\altaffilmark{1,2},
Puragra~Guhathakurta\altaffilmark{3} 
Rachael~L.~Beaton\altaffilmark{4},
James~Bullock\altaffilmark{5},
Marla~C.~Geha\altaffilmark{6},
Jason~S.~Kalirai\altaffilmark{7,8},
Evan~N.~Kirby\altaffilmark{9,2},
Steven~R.~Majewski\altaffilmark{4},
James~C.~Ostheimer\altaffilmark{4},
Richard~J.~Patterson\altaffilmark{4},
Erik~J.~Tollerud\altaffilmark{5},
Mikito~Tanaka\altaffilmark{10}, and
Masashi~Chiba\altaffilmark{10}
}

\email{
kgilbert@astro.washington.edu}

\altaffiltext{1}{Department of Astronomy, University of Washington, Box 351580, Seattle, WA 98195-1580, USA}
\altaffiltext{2}{Hubble Fellow.}
\altaffiltext{3}{UCO/Lick Observatory, Department of Astronomy \&
Astrophysics, University of California Santa Cruz, 1156 High Street, Santa
Cruz, CA 95064, USA}
\altaffiltext{4}{Department of Astronomy, University of Virginia, PO Box~400325, Charlottesville, VA 22904-4325, USA}
\altaffiltext{5}{Center for Cosmology, Department of Physics and Astronomy, University of California at Irvine, Irvine, CA, 92697, USA}
\altaffiltext{6}{Astronomy Department, Yale University, New Haven, CT 06520, USA}
\altaffiltext{7}{Space Telescope Science Institute, Baltimore, MD 21218, USA}
\altaffiltext{8}{Center for Astrophysical Sciences, Johns Hopkins University, Baltimore, MD, 21218, USA}
\altaffiltext{9}{California Institute of Technology, 1200 East California Boulevard, MC 249-17, Pasadena, CA 91125, USA}
\altaffiltext{10}{Astronomical Institute, Tohoku University, Aoba-ku, Sendai 980-8578, Japan}

\setcounter{footnote}{11}

\begin{abstract}
We present the surface brightness profile of M31's stellar halo out to 
a projected radius of 175~kpc. 
The surface brightness estimates are based on 
confirmed samples of M31 red giant branch stars derived from Keck/DEIMOS spectroscopic
observations.  
A set of empirical spectroscopic and photometric M31 membership diagnostics is used 
to identify and reject foreground and background contaminants. 
This enables us to trace the stellar halo of M31 to larger projected distances and 
fainter surface brightnesses than previous photometric studies.  
The surface brightness profile of M31's 
halo follows a power-law with index $-2.2\pm0.2$ and extends to a projected distance of at least 
$\sim 175$~kpc ($\sim 2/3$ of M31's virial radius), with no evidence of a downward break at large radii. 
The best-fit elliptical isophotes
have $b/a=0.94$ with the major axis of the halo aligned along the minor axis of M31's disk, 
consistent with a prolate halo, although 
the data are also consistent with M31's halo
having spherical symmetry. 
The fact that tidal debris features are kinematically cold is used to 
identify substructure in the spectroscopic fields out to projected radii 
of 90~kpc, and investigate the effect of this substructure
on the surface brightness profile.   The scatter in the surface 
brightness profile is reduced when kinematically 
identified tidal debris features in M31 are 
statistically subtracted; the remaining profile indicates that a comparatively 
diffuse stellar component to M31's stellar halo exists to large distances.  Beyond
90~kpc, kinematically cold tidal debris features can not be identified due to small
number statistics; nevertheless, the significant field-to-field variation in surface brightness
beyond 90~kpc suggests that the outermost region of M31's halo is also comprised to 
a significant degree of stars stripped from accreted objects. 
\end{abstract}


\setcounter{footnote}{0}

\section{Introduction}\label{sec:intro}

Stellar halos of galaxies are low density environments in which the 
detritus of hierarchical structure formation can remain visible for Gigayears
in the form of tidal debris structures.  Detailed simulations of 
stellar halo formation in a cosmological context 
\citep{bullock2001,bullock2005,font2006,font2008,johnston2008,zolotov2009,zolotov2010,cooper2010,font2011} 
are providing a growing framework for interpreting 
observations of tidal debris in stellar halos.  
Current simulations suggest that galaxy mergers play a primary
role in the formation of stellar halos: the outer regions of 
stellar halos are built via the
accretion of smaller galaxies and tidal stripping of their stars,
while the inner regions may be built through a combination of
merging events and in situ star formation \citep{zolotov2010,font2011,mccarthy2012}. 
Thus, the global properties of stellar halos are a product 
of the merging history of the host galaxy.   

Recent observational campaigns have extended studies of stellar halos 
beyond the Local Group \citep{tanaka2011,martinez-delgado2010,radburn-smith2011}. 
However, M31 remains one of the best laboratories for studying stellar
halos.  M31's proximity enables us to study its resolved stellar 
population not just with photometry, but also through spectroscopy of 
individual stars.  

Early work on the integrated 
light of M31 \citep{de-vaucouleurs1958,walterbos1987} was 
followed by resolved stellar population studies over the last several decades 
\citep[e.g.,][]{mould1986,ferguson2002}.  
These studies found that within $R\sim30$~kpc, the M31 stellar halo appeared to be a continuation of M31's inner bulge, and was characterized by the following properties: a de~Vaucouleurs $r^{1/4}$ law radial surface brightness profile \citep{pritchet1994}, a high stellar density \citep{reitzel1998}, and a high mean metallicity \citep[$\rm{[Fe/H]}\sim -0.5$][]{mould1986,rich1996,durrell2001,reitzel2002,bellazzini2003,durrell2004}.
Furthermore, deep imaging (obtained with the Advanced Camera for Surveys on the 
{\it Hubble Space Telescope}) 
along M31's minor axis at radii of 10\,--\,35 kpc (well beyond
the extent of M31's disk)
revealed a significant population of $\sim6$\,--\,8~Gyr stars
\citep{brown2003,brown2006,brown2007,brown2008}. 
The properties of the inner regions of M31's spheroid stand 
in stark contrast to the properties of the Milky Way's (MW) stellar 
halo, which consists almost entirely of old, metal-poor stars and appears 
to follow a power-law density profile ($\rho(r)\propto r^\gamma$) .  
Studies of the MW's halo have typically measured $\gamma \sim -2.7$ to $-3.5$ \citep[e.g.,][]{morrison2000,yanny2000,siegel2002,juric2008,sesar2011}, 
equivalent to a surface density distribution with a slope of -1.7 to -2.5.
 
In the past five years, large photometric and spectroscopic surveys 
have discovered and begun to characterize M31's outer stellar halo,
which has a power-law surface brightness profile, 
low stellar density, and metal-poor stars. 
Using the spectroscopic data set of \citet{gilbert2006}, 
\citet{guhathakurta2005} showed that beyond projected radial distances 
of \rproj\,$\sim20$\,--\,30~kpc, 
the surface brightness profile of M31's stellar halo is consistent 
with a power-law of index $-2.3$ out to the
limits of the surveyed region, \rproj\,$\sim 165$~kpc.  
A concurrent photometric study based on 
data from the Isaac Newton Telescope Wide-Field Camera 
\citep{irwin2005} also found evidence for a break in the 
radial surface brightness profile of M31 at \rproj\,$\sim 20$~kpc, and 
determined that beyond this radius the profile flattens and is consistent 
with a power-law index of $-2.3$; their study reached to a projected 
radial distance of 55~kpc from the center of M31.  
\citet{ibata2007} presented a large photometric survey of M31's 
southern quadrant undertaken with the Canada--France--Hawaii 
Telescope/MegaCam and found that the radial surface brightness profile of M31 out 
to \rproj\,$\sim 130$~kpc is consistent with a power-law of index $-1.9$.

M31's outer stellar halo consists of stars that
are on average more metal-poor than the stars that comprise the inner,
bulge-like spheroid.  
\citet{kalirai2006halo} used the data set from \citet{gilbert2006}
to analyze the metallicity of 
red giant branch (RGB) stars from \rproj\,$\sim 11$ 
to 165~kpc in M31's southern quadrant.  They found that while 
the inner regions of M31's
spheroid are relatively metal-rich, the stellar population becomes
increasingly metal-poor beyond \rproj\,$\sim30$~kpc 
\citep[see also][]{koch2008,tanaka2010}.  A contemporaneous study 
found evidence of metal-poor M31 RGB stars from \rproj\,$\sim10$ to 
70~kpc in fields located near M31's major axis \citep{chapman2006}.

Most previous studies of M31's stellar halo consisted primarily of fields 
in the southern quadrant of M31 or along M31's southern minor axis. 
Two recent photometric surveys have extended observations of M31's stellar 
halo into the other three quadrants, providing a more global view 
of the halo.  The Pan-Andromeda Archaeological Survey 
\citep[PAndAS;][]{mcconnachie2009} used the Canada--France--Hawaii Telescope (CFHT)
and the MegaCam instrument to survey 
the eastern, northern, and western quadrants of M31's halo
out to a maximum projected radial distance of $\sim 150$~kpc.  PAndAs found stellar sources 
consistent in color and apparent magnitude with RGB stars at the distance of M31 
throughout the survey volume \citep{mcconnachie2009}.  \citet{tanaka2010} surveyed
along M31's minor axis with the 
Suprime-Cam instrument on the Subaru Telescope, achieving a deeper 
photometric limit than 
PAndAS, although over a much more limited area.  They observed strips of 
contiguous fields on the 
southeast and northwest minor axis out to projected
distances of \rproj\,$\sim 60$ and $\sim 100$~kpc, respectively, and found
the radial surface brightness profile along M31's northwest minor axis is
described by a power-law with an index of $-2.2$, 
consistent with results from the southeast minor axis. 

Interpretation of the global properties of M31's stellar halo is 
complicated by the presence of large, extended, and numerous
tidal debris features
 \citep[e.g.,][]{ibata2001nature,ferguson2002,kalirai2006gss,ibata2007,gilbert2007,gilbert2009gss,mcconnachie2009,tanaka2010}. 
It is difficult to assess the effect of substructure on measured values 
(mean metallicity, surface brightness profile) with purely photometric 
studies.  The surface brightness profile of the 
smooth underlying stellar population must be determined 
by removing entire areas with photometrically identified substructure.  In much of 
M31's stellar halo, this requires removing a significant fraction of the 
available spatial area from the analysis \citep[e.g.,][]{ibata2007,tanaka2010}. 
Spectroscopy of the stellar population yields the kinematical information needed 
to separate intact, dynamically
cold tidal debris features (typically with $\sigma_v\lesssim30$~\kms)
from an underlying spatially diffuse, dynamically hot population 
($\sigma_v\gtrsim100$~\kms).

This paper is the first in a series that describes
the global properties of M31's stellar halo using spectroscopic 
observations of member RGB stars in all quadrants of 
M31's stellar halo.  
In this contribution, we measure the surface brightness profile of M31's
stellar halo using counts of spectroscopically confirmed M31 RGB stars and  
quantify the effect of substructure on the surface brightness profile. 
The profile presented here encompasses data from more 
than 100 spectroscopic 4\arcmin$\times$16\arcmin\ slit masks in 38 fields; 27 fields are being
presented for the first time.  The fields cover a range of position 
angles and projected distances from M31, and were all obtained with the DEIMOS 
spectrograph on the Keck telescope.  
The data included in our analysis are discussed in 
Section~\ref{sec:data}.  The method used to convert counts of spectroscopically 
identified M31 RGB stars to a surface brightness estimate is described in 
Section~\ref{sec:sb_est}.  The 
global surface brightness profile of M31's stellar halo is presented
and analyzed in Section~\ref{sec:sbprofile}.  The results are discussed within
the broader context of hierarchical stellar halo formation in Section~\ref{sec:discussion}.
Finally, we summarize our conclusions in Section~\ref{sec:conclusion}.
A distance modulus of 24.47 is assumed for all conversions of
angular to physical units 
\citep[corresponding to a distance to M31 of 783~kpc;][]{stanek1998}.

\section{Data}\label{sec:data}

\begin{figure*}[tb!]
\plotone{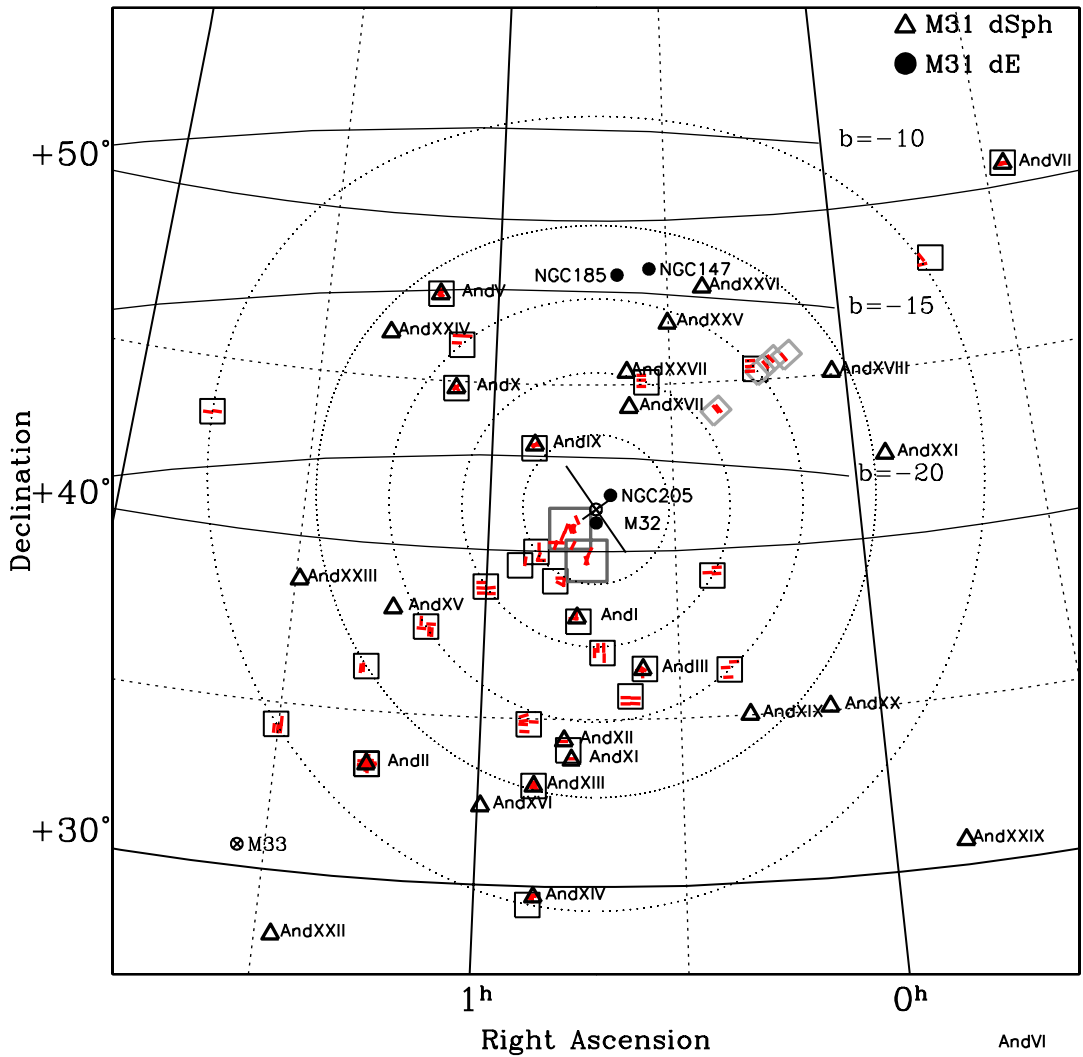}
\caption{Locations and orientations of the spectroscopic masks 
(red rectangles) included in the analysis (Section~\ref{sec:data}).  
The footprints of the KPNO/Mosaic (black), CFHT/MegaCam (dark gray), and Subaru/Suprime-Cam (light gray)
images used to design the spectroscopic slit masks are also shown.  The center of M31 
is marked with an open circle, and the orientations of the major and minor axes of M31's 
disk are denoted respectively by the long and short solid lines.  The dwarf elliptical 
satellites of M31 are denoted by circles, and the dwarf spheroidal (dSph) satellites of M31 
are denoted by triangles.
The dotted circles denote radii of 2, 4, 6, 8, and 11 deg from M31's center. 
}
\label{fig:roadmap}
\end{figure*}

We present data in 38 fields spanning a large range in position angle and 
projected distance from the center of M31 
(Figure~\ref{fig:roadmap} and Table~\ref{tab:fields}). Thirteen fields lie outside the
southern quadrant (the most heavily observed quadrant of M31's halo), and 
10 fields lie at projected distances of 100 kpc or more from M31's center.    
The data were obtained as part of the SPLASH 
(Spectroscopic and Photometric Landscape of Andromeda's Stellar Halo) survey
of M31's stellar halo.  In this section, we summarize the photometric and 
spectroscopic observations and data reduction (Section~\ref{sec:obs}) and 
the identification of the M31 RGB star sample 
(Sections~\ref{sec:cleansample} and \ref{sec:dsph_fields}).  
Further details of the observations and reduction techniques
used in the SPLASH survey can be found in the listed references.

\subsection{Photometric and Spectroscopic Observations}\label{sec:obs}

The photometric data are primarily from imaging observations 
obtained with the Mosaic Camera on the Kitt Peak National Observatory (KPNO) 
4-m Mayall telescope.\footnote{Kitt Peak National
Observatory of the National Optical Astronomy Observatory is operated by the
Association of Universities for Research in Astronomy, Inc., under cooperative
agreement with the National Science Foundation.}
Photometric catalogs were derived from observations in the Washington 
system $M$ and $T_2$ filters and the intermediate-width DDO51 filter 
\citep{ostheimer2003}.  The observed magnitudes were transformed
to $V$ and $I$ using the equations of \citet{majewski2000}.  

Additional photometric data in the outer (\rproj\,$>30$~kpc) fields came
from two sources.  The photometry in field `d10' was derived from 
$V$ and $I$ images obtained with the William Herschel Telescope \citep{zucker2007}.  
Photometry for fields `streamE' and `streamF' was derived from  $V$ and $I$
images obtained with the Suprime-Cam instrument on the Subaru Telescope \citep{tanaka2010}.

Photometric catalogs for the innermost fields (\rproj\,$<30$~kpc) 
were derived from observations obtained with the MegaCam instrument on 
the 3.6-m CFHT.\footnote{MegaPrime/MegaCam 
is a joint project of CFHT and CEA/DAPNIA, 
at the Canada-France-Hawaii Telescope
which is operated by the National Research Council of Canada, the Institut
National des Science de l'Univers of the Centre National de la Recherche
Scientifique of France, and the University of Hawaii.} 
Images were obtained with the $g'$ and $i'$ filters.  
The observed stellar magnitudes were transformed to Johnson-Cousins 
$V$ and $I$ magnitudes
using observations of Landolt photometric standard stars as described in 
\citet{kalirai2006halo}.

The above photometric catalogs were used to design spectroscopic 
slit masks for use in the DEIMOS spectrograph on the Keck~II telescope.  
Table~\ref{tab:fields} lists the filters used for each of the 
spectroscopic fields.  Objects were assigned a priority for inclusion on the
slit mask based on image morphology, their
$I_0$ magnitude, and (when available) the probability of
their being RGB stars at the distance of M31 
based on their position in an ($M-T_2$, $M-$DDO51) color--color diagram 
\citep{palma2003,majewski2005}.  The mask design process is 
described in detail by \citet{guhathakurta2006}. 
In the high-density inner 
regions of M31's spheroid not all high priority M31 RGB candidates within the 
DEIMOS mask area can be included due to slit conflicts.  In contrast, 
the low density outer fields contain a high fraction of low priority filler 
targets due to the paucity of M31 RGB candidates.  The effect of this
radial dependence of slit mask targets on the surface brightness estimates 
is discussed in Section~\ref{sec:lumfun}.

Spectra were obtained over nine observing seasons (Fall 2002\,--\,2010) with
the DEIMOS spectrograph on the Keck~II 10-m telescope.  
The 1200~line~mm$^{-1}$ grating, which has a dispersion of $\rm0.33~\AA$~pixel$^{-1}$, 
was used for all observations.  
The slit width of 1\arcsec\ yields a resolution of 1.6~\AA\ FWHM.
The typical wavelength range of the spectral 
observations is 6450\,--\,9150\AA.   This wavelength range includes the \caii\ 
triplet absorption feature at $\sim8500$\AA\ and the \nai\ absorption feature at 
8190\AA.  Masks were observed for approximately 1 hr each, with modifications
made for particularly good or bad observing conditions.   

The spectra were reduced using modified 
versions of the {\tt spec2d} and {\tt spec1d} software developed 
at the University of California, Berkeley \citep{newman2012,cooper2012}. 
The {\tt spec2d} routine is used to perform the flat-fielding, 
night-sky emission line removal and extraction of one-dimensional spectra
from the two-dimensional spectral data. 
The {\tt spec1d} routine cross-correlates the resulting 
one-dimensional spectra with template spectra to determine the redshift
of the objects; the template library includes
stellar spectra obtained with DEIMOS and galaxy spectra from the Sloan 
Digital Sky Survey \citep[see ][for details]{simon2007}. 
Each spectrum was visually inspected, and only stellar spectra with secure 
velocity measurements were included in the analyzed data set \citep{gilbert2006}.  
Two corrections were applied to the observed velocities: (1) a heliocentric correction,
and (2) a correction for imperfect centering of 
the star within the slit.  The latter correction was calculated by measuring 
the observed position of the atmospheric $A$-band absorption feature 
relative to night sky emission lines \citep{simon2007,sohn2007}.  

We are presenting data from 108 spectroscopic masks.   These masks tareted
$\sim 11000$ objects, and yielded successful velocity measurements from over 5800 stellar spectra (52.2\%).  The remainder of the targets were galaxy spectra (20.2\%), spectra for which a velocity measurement was not possible due to insufficient signal-to-noise ratio (S/N) or a lack of strong spectral lines (24.2\%), and a small fraction of catastrophic failures (3.4\%).

\subsection{Classification of M31 RGB Stars and Foreground MW Stars }\label{sec:cleansample}
\begin{figure*}[htb!]
\centerline{
\includegraphics[width=0.5\textwidth]{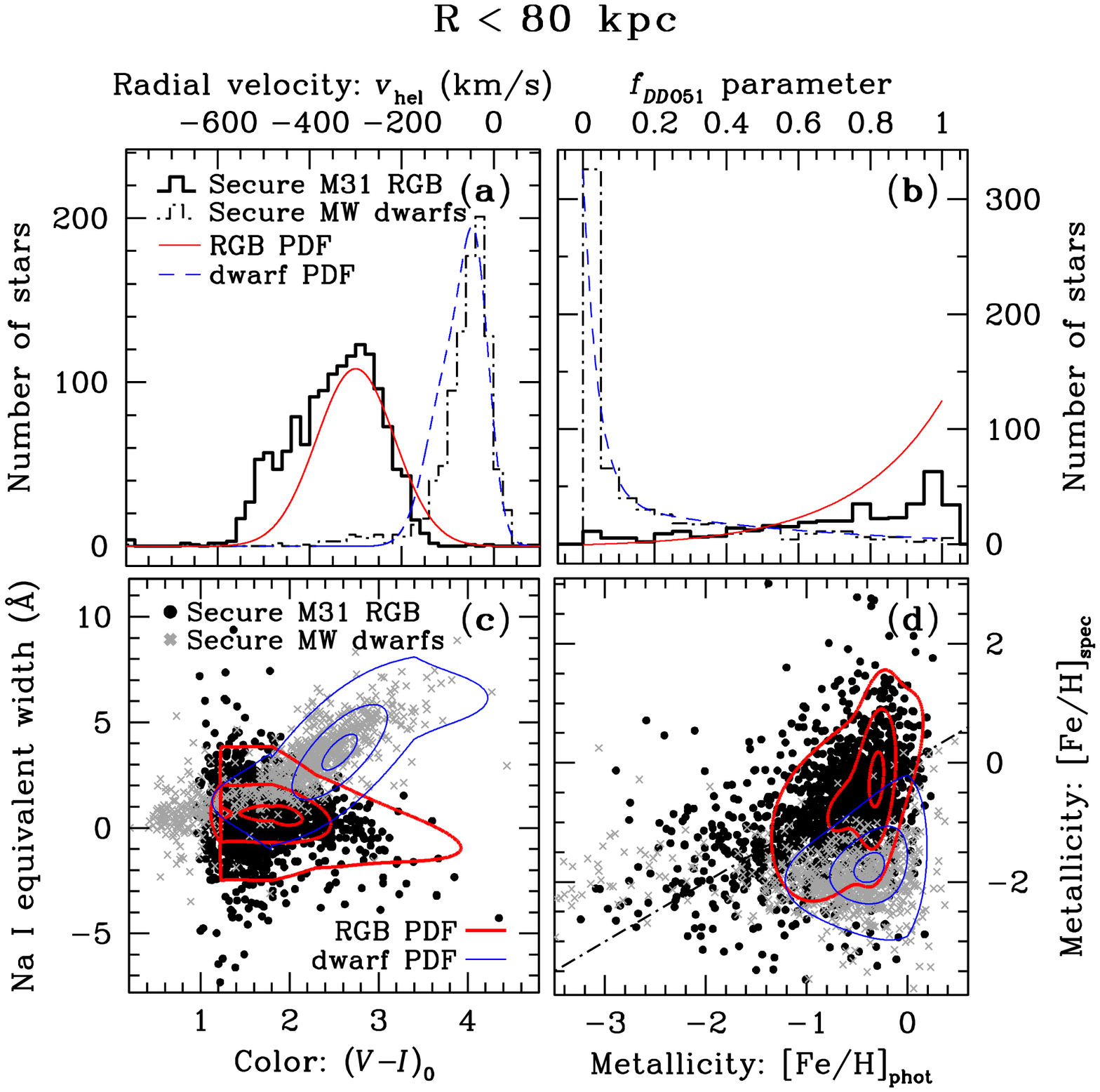}
\includegraphics[width=0.5\textwidth]{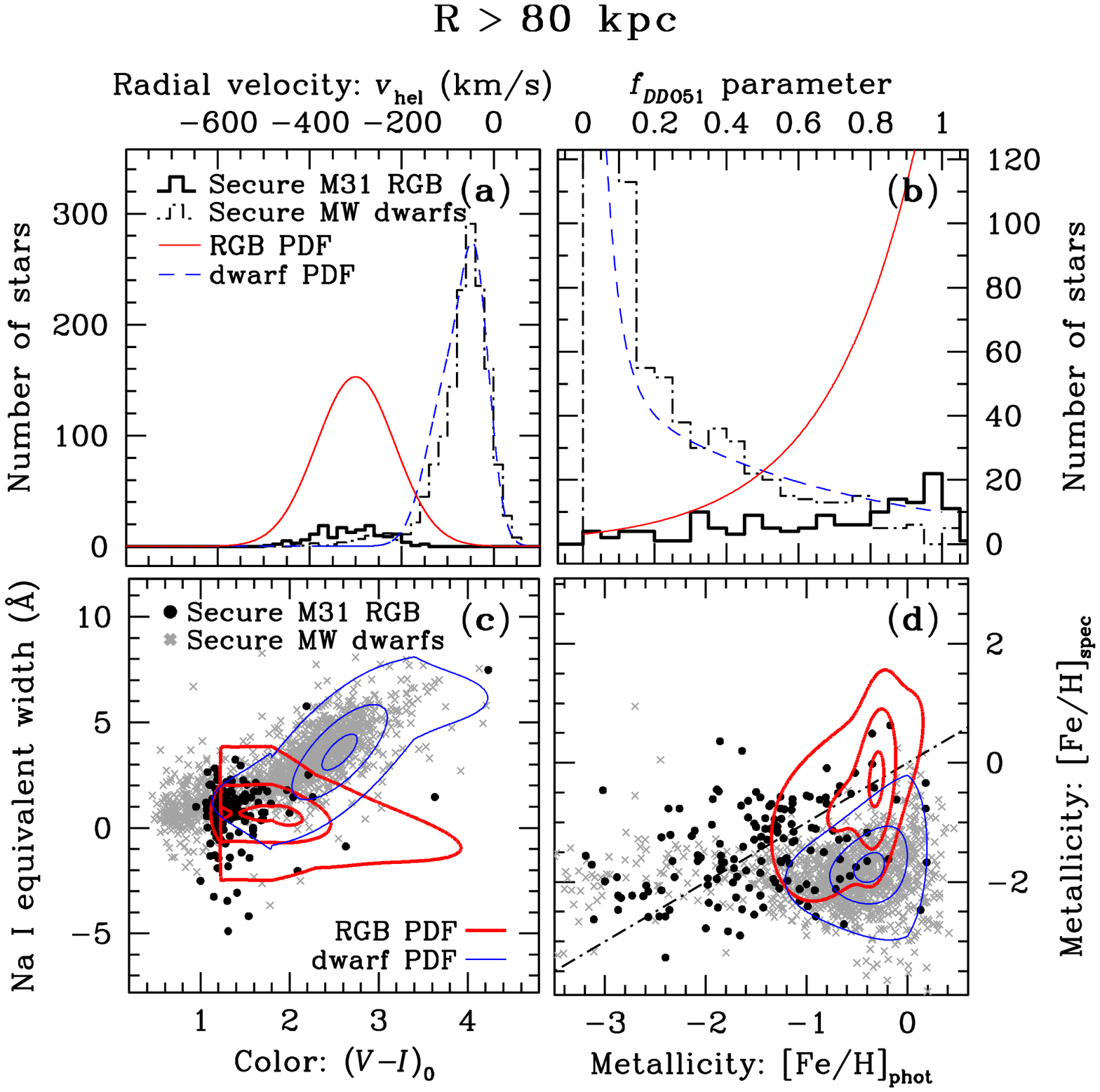}
}
\caption{Locations of stars in four of the five empirical diagnostics \citep[Section~\ref{sec:cleansample};][]{gilbert2006}: (a) line-of-sight velocity, (b) DDO51 parameter (Section~\ref{sec:obs}), (c) \nai\ EW, and (d) spectroscopic versus photometric [Fe/H] estimates.  
Left 4 panels: stars identified as secure M31 red giants (classes +2 and +3) and MW dwarfs (classes $-2$ and $-3$) in fields from \rproj\,$\sim 8$ to 80~kpc.  The majority of the stars come from the interior fields (\rproj\,$<30$~kpc).
Right 4 panels: stars identified as secure M31 red giants and MW dwarfs in the outer fields (\rproj\,$\gtrsim 80$~kpc).  The RGB and dwarf probability distribution functions shown in panels (a) and (b) are normalized to equal area. 
The PDF contours in panels (c) and (d) show the 90\%, 50\%, and 10\% contours.
Although significantly fewer in number and primarily metal-poor (as can be seen by comparing panel (d) in the left and right figures), stars identified as M31 RGB stars in the outer fields are consistent with the distribution of 
M31 RGB stars in the inner fields. 
}
\label{fig:global_diags}
\end{figure*}

MW dwarf stars along the line of sight constitute a significant fraction of 
the observed stellar spectra even 
though the sample of spectroscopic targets consists primarily of objects that have been photometrically preselected to be M31 RGB candidates.
Moreover, the line-of-sight velocity
distributions of M31 halo RGB and MW dwarf stars overlap, and the 
velocity distribution of foreground MW stars has a tail that extends well into
the velocity range typical of M31 halo stars. 
To estimate the surface
brightness of a given M31 field, the relative fractions
of stars in each of the two populations must be securely measured.

Individual stars are identified as M31 red giants or MW dwarfs using the method 
developed by \citet{gilbert2006}, to which readers are referred for full details
of the technique.  A combination of photometric and 
spectroscopic diagnostics are used to determine the probability an individual star is 
an M31 red giant or MW dwarf: (1) line-of-sight velocity ($v_{\rm los}$), 
(2) photometric probability
of being a red giant based on location in the ($M-T_2$, $M-$DDO51) color-color 
diagram (when available; Table~\ref{tab:fields}), (3) the equivalent width of the \nai\ 
absorption line (surface-gravity and temperature sensitive) versus \vio\ color, 
(4) position in the \ivi\ CMD and (5) spectroscopic (based on the EW of the 
\caii\ triplet) versus photometric (comparison to theoretical RGB 
isochrones) \feh\ 
estimates.  Each diagnostic provides separation between M31 RGB stars and MW 
dwarf stars based on different physical parameters.  A star's location in 
each diagnostic is compared
to empirical probability distribution functions (PDFs) to determine the likelihood 
($L_i={\rm log}_{10}(P_{\rm RGB}/P_{\rm dwarf}$)) the
star is an M31 red giant or MW dwarf (Figure~\ref{fig:global_diags}); each PDF was determined
using training sets of M31 red giant and MW dwarf stars \citep{gilbert2006}.   
The likelihoods for each diagnostic are
combined to give the overall likelihood, \olkhd, the star is an M31 red giant or MW dwarf. 
Stars that are $\geq 3$~times more probable to be an M31 red giant than MW dwarf star 
(\olkhd\,$\geq 0.5$)
are designated as secure M31 red giants (and vice versa for secure MW dwarf stars: \olkhd\,$\leq -0.5$); 
stars below this probability ratio threshold are designated as marginal M31 red giants ($0<$\olkhd\,$< 0.5$) or MW dwarfs ($-0.5<$\olkhd\,$\le 0$).
The analysis is restricted to stars designated as secure M31 red giants or MW dwarfs because
the secure categories suffer from the least contamination \citep{gilbert2007}.  

Figure~\ref{fig:global_diags} presents the position in the four most 
powerful diagnostics of secure M31 RGB and MW dwarf stars 
in fields interior (left panels) 
and exterior (right panels) to \rproj\,$\sim 80$~kpc.   
Very few stars are identified as secure M31 RGB stars in the outermost fields, 
however Figure~\ref{fig:global_diags} shows that 
those that are identified as M31 red giants are more consistent with the M31 RGB, 
rather than the MW dwarf, PDFs.
The RGB stars in the outer
fields (\rproj\,$\gtrsim 80$~kpc) are bluer (more metal-poor) than the 
average RGB star in the inner fields, as expected based on the observed metallicity gradient
of M31's stellar halo \citep{kalirai2006halo,chapman2006,koch2008}.
However, their locations in the four diagnostics are consistent with the distribution of 
M31 RGB stars in the inner fields.  

The noticeable asymmetry in the velocity distribution of M31 RGB stars within 80~kpc
(Figure~\ref{fig:global_diags}(a), left panel) is due to two effects.  
The first is the presence of stars associated with the giant southern stream 
in many of the south quadrant M31 halo fields.  The velocity of the giant 
southern stream varies 
with \rproj, but remains more negative than M31's systemic velocity, 
$v_{\rm sys}^{\rm M31}=-300$~\kms\ 
\citep{ibata2005,guhathakurta2006,kalirai2006gss,gilbert2009gss}.  
The second effect is incompleteness in the secure sample of M31 RGB stars 
at the positive end of the M31 velocity distribution.  The velocity 
diagnostic typically provides greater separation of the populations, 
and thus has intrinsically higher weight, than the other diagnostics.
Thus the velocity region where the M31 and MW velocity distributions
overlap tends to have a higher percentage of stars identified as marginal 
M31 RGB stars \citep[for a full discussion, see the appendix of][]{gilbert2007}.

\subsubsection{Comparison of M31 RGB and MW Dwarf Star Samples}\label{sec:m31mwsamples}
\begin{figure*}[htb!]
\plotone{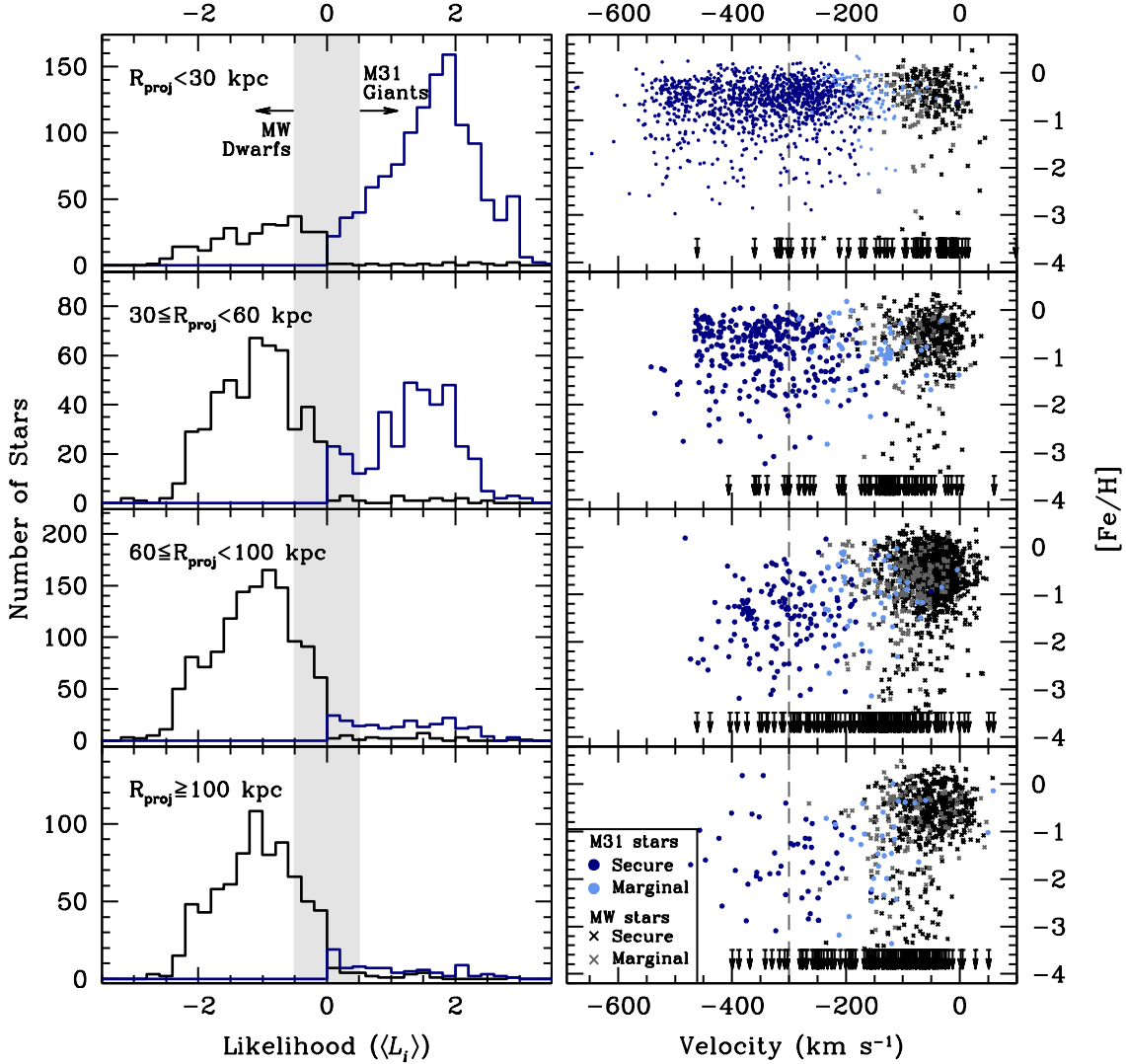}
\caption{Left: likelihood distribution (Section~\ref{sec:cleansample}) of all observed stars as a function of projected distance from M31's center.  Blue histograms show the distribution of stars classified as M31 RGB stars, while black histograms denote foreground MW dwarf stars.  The shaded area shows the region of likelihood space excluded from the analysis: stars that have likelihood values in this range are considered marginal identifications.  Some stars are classified as MW dwarf stars even though \olkhd\,$\ge0$.  These are stars whose $V_0, I_0$ magnitudes place them bluer than the most metal-poor isochrone by an amount greater than the photometric error (leading to an extrapolated \fehp $\lesssim -3.5$, see right panels).  Right: distribution of stars in the \fehp--$v_{\rm los}$ plane.  Marginal stars are those with $-0.5\le$\olkhd\,$\le0.5$.  Although the relative fraction of M31 RGB stars and MW dwarf stars changes with radius, the locus of M31 RGB stars in the \fehp--$v_{\rm los}$ plane remains populated even at large projected distances (\rproj\,$\ge 100$~kpc) from M31's center.  M31's systemic velocity is $v_{\rm los}=-300$~\kms.   
}
\label{fig:likelihoods}
\end{figure*}

\begin{figure}[htb!]
\plotone{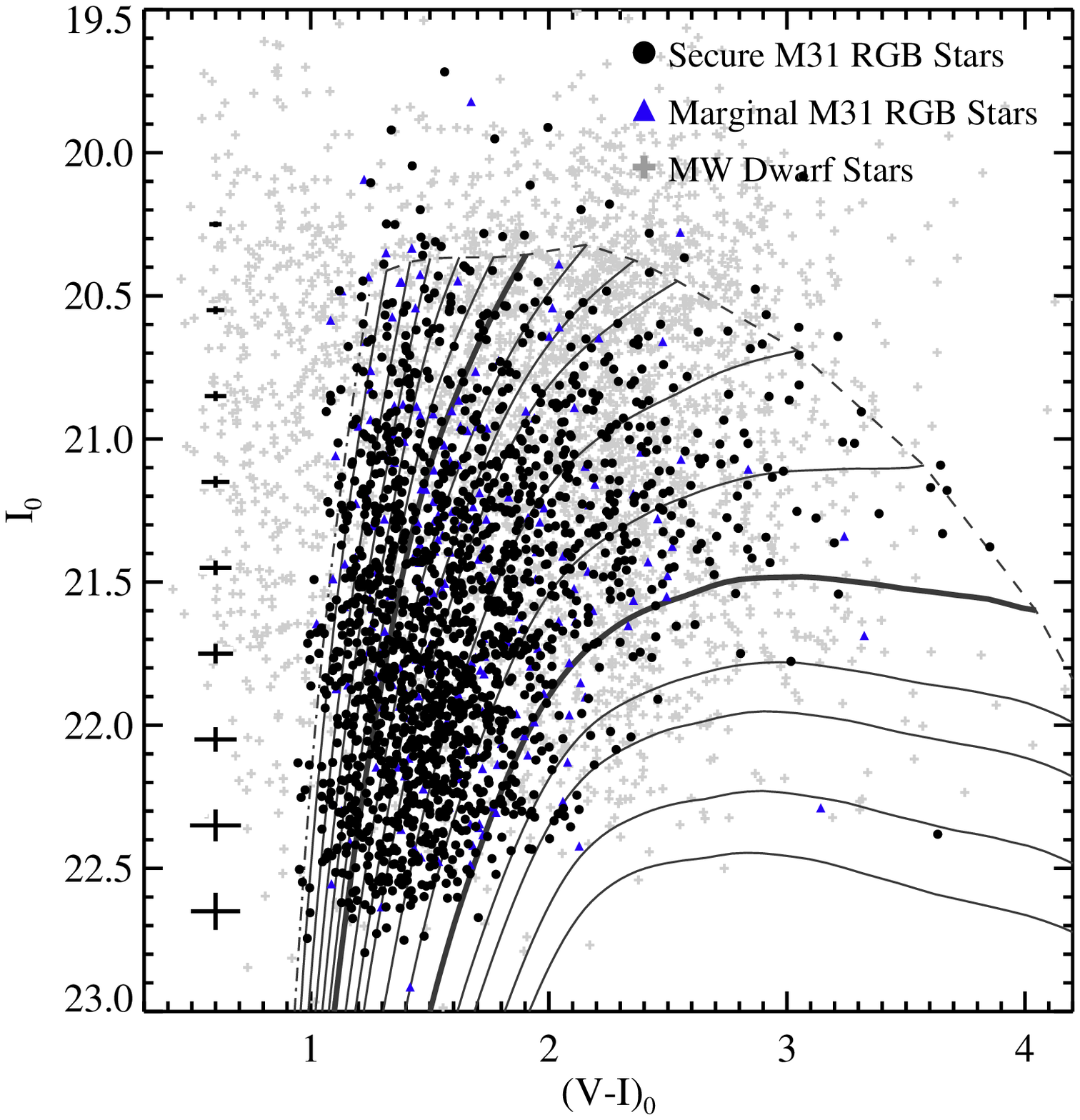}
\caption{$I_0, (V-I)_0$ color--magnitude diagram for the spectroscopic sample, with theoretical isochrones of \citet{vandenberg2006} ranging from [Fe/H]=$-2.3$ (left most) to $0.49$ (right most) overlaid (assuming an age of 10~Gyr and [$\alpha$/Fe]=0.0).  The thick isochrones denote metallicities of [Fe/H]=$-1.0$ and $0.0$, while the dashed isochrone is for an age of 5~Gyr and [Fe/H]=$-2.3$.  Stars classified as red giants at the distance of M31 trace the RGB isochrones, while stars classified as MW dwarf stars along the line of sight have a broader color distribution.  Typical photometric errors as a function of $I$ magnitude are shown on the left: stars that are bluer than the most metal-poor isochrone by more than the typical photometric error are classified as MW dwarf stars (Section~\ref{sec:m31mwsamples}). 
}
\label{fig:cmd}
\end{figure}

The left panels of Figure~\ref{fig:likelihoods} show the distribution of stars in likelihood 
space.  As discussed above, stars 
with marginal M31/MW classification (shaded region of left panels) tend to 
lie in the region of velocity space where the RGB and MW dwarf star PDFs 
overlap.  Although the sample is divided into M31 and MW stars at \olkhd=0, 
the true likelihood distributions of M31 (\olkhd$>0$) and MW stars (\olkhd$<0$) 
are expected to have tails that extend into negative and positive 
likelihood values, respectively.  Restricting the sample to secure M31 (\olkhd$\ge0.5$) and 
secure MW (\olkhd$\le-0.5$) stars minimizes the contamination in 
the M31 and MW samples.  

The right panels of Figure~\ref{fig:likelihoods} display the location of stars in
the plane defined by heliocentric line-of-sight velocity and photometric metallicity (\feh). 
\feh\ values are determined by comparison of each star's $I_0$ magnitude and $(V-I)_0$ 
color with a grid of theoretical RGB isochrones at the distance of M31 (Figure~\ref{fig:cmd}).  
For stars that are either slightly bluer than the bluest isochrone, or slightly 
above the TRGB, the metallicities are measured by extrapolating the grid, 
enabling a continuous mapping of the color-magnitude distribution of stars 
to \fehp\ space.  Clearly, these isochrone based \fehp\ estimates
are only physically meaningful for the bona fide M31 RGB stars.
Regardless, in the \feh--$v_{\rm los}$ plane, three distinct groups of stars 
are apparent in the sample.  As discussed below, these groups represent 
RGB stars in the halo of M31, dwarf 
stars in the MW disk, and main-sequence turn off stars in the MW 
halo.   

The M31 RGB star locus is centered at $v_{\rm los}=-300$~\kms\ (the 
systemic velocity of M31) and 
spans a reasonable range of \fehp\ for RGB stars.  
Even in the most distant fields (\rproj\,$\ge 100$~kpc), 
the sample includes stars that are clearly located in the M31 locus.

The two MW dwarf star loci denote the two primary populations of MW 
contaminants expected in the survey. The MW dwarf locus at velocities 
$-150\lesssim v_{\rm los}\lesssim 50$~\kms\ and relatively red colors 
(which translate into high \fehp\ estimates) is consistent with the properties
expected for stars 
belonging to the disk of the MW.   

The second MW dwarf locus has a 
wide spread in velocities 
($-400\lesssim v_{\rm los}\lesssim 0$) and significantly bluer colors 
than either the M31 RGB stars or MW disk stars; indeed, the $(V-I)_0$ colors
of these stars place them well to the blue of the RGB fiducials, 
and thus these stars are depicted with upper limits.
This locus is consistent with the properties expected for main sequence 
turn-off stars at varying line-of-sight distances in the {\it Galactic} halo. 
The stars with \olkhd$>0$ in Figure~\ref{fig:likelihoods} that are
classified as MW stars are part of this population.
These stars'  velocities ($v_{\rm los}\lesssim -200$~\kms), combined with their 
blue colors [\vio$\lesssim 1$], typically lead to high (\olkhd$>0$) likelihood 
values because the CMD, \nai\ EW, and [Fe/H] diagnostics have no discriminating 
power at these colors (Figure~\ref{fig:global_diags}).  Therefore, 
any star that is blue-ward of the most metal-poor isochrone by more 
than the typical photometric error is classified as an MW dwarf star 
\citep[Figure~\ref{fig:cmd}; see Section~4.1.2 of][for a more detailed discussion]{gilbert2006}.  
It is possible that a small fraction of these
could be M31 stars.  Individual stars might have spurious photometric measurements, 
or they could be asymptotic giant 
branch stars or very metal-poor M31 halo red giants.  However, 
the vast majority of these stars do 
appear to lie well-removed from the locus of M31 stars and 
to be part of a separate population of stars, with properties consistent 
with main sequence turn-off stars in the Galactic halo (Figure~\ref{fig:likelihoods}).

\subsection{Identification of M31 Field Stars in Masks Targeting dSph Galaxies}\label{sec:dsph_fields}
The SPLASH survey includes an effort to measure the kinematical and 
chemical properties of M31's dwarf satellite galaxies using spectroscopy of
member stars \citep{kalirai2009,kalirai2010,tollerud2012}.  
Thus, a number of the M31 spectroscopic fields presented here 
were designed to target dSphs in the halo of M31 (Table~\ref{tab:fields}). 
The number of field stars belonging to M31's stellar halo in 
the dSph masks is estimated following the technique used by \citet{gilbert2009gss} 
(their Section~2.5.1, Figures 3 and 4).
The distribution of stellar objects in velocity, metallicity, and sky
position is used to isolate RGB stars that are likely part of M31's stellar
halo rather than members of the dwarf satellite.  To classify the stars, 
we leverage the following three properties of the dSph galaxies: 
(1) they are generally compact enough to cover only a portion of the 
spectroscopic slit mask, 
(2) they have small velocity dispersions ($< 10$~\kms), and 
(3) their member stars generally span a limited range of [Fe/H].  
All RGB stars beyond the estimated King limiting radius of the dSph \citep{mcconnachie2006,martin2006,majewski2007,zucker2007,collins2010} 
are designated as 
M31 field stars.  
RGB stars that are inside the limiting radius of the dSph but well removed 
from the tight locus in $v_{\rm los}$\,--\,[Fe/H] space occupied by the dSph members 
are also designated as M31 field stars.  
Table~\ref{tab:fields} lists literature references
for each dSph, where the distribution of stars and their membership is 
discussed in detail. 

Although this procedure will miss M31 field stars that happen to fall both 
within the limiting radius of the dSph galaxy and in the region of 
$v_{\rm los}$\,--\,[Fe/H] space occupied by the dSph members, the number of 
such M31 field stars is expected to be small.  This is due in part to the fact
that in each field, the dSph members span a very narrow range in velocity space 
compared to M31 halo stars.  It is also due to the fact that the surface 
brightness of each dSph becomes increasingly dominant over the surface 
brightness of M31's spheroid toward the center of the dSph, resulting in an 
increased likelihood that spectroscopic slits will be placed on a dSph member 
rather than an M31 field star. 

\section{Surface Brightness Estimates from M31 RGB Star Counts}\label{sec:sb_est}
The surface brightness of M31's stellar halo is estimated from counts of 
spectroscopically confirmed M31 RGB stars (Section~\ref{sec:cleansample}).  
The ratio of securely identified RGB stars ($N_{\rm M31}$) to securely identified MW dwarf 
stars ($N_{\rm MW}$) in a field is used to estimate M31's stellar surface density.  
However, the density of foreground MW dwarf stars is not constant over the widely spaced spectroscopic fields.  To account for this, the ratio of observed M31 RGB and MW dwarf stars is multiplied by the total expected surface density of MW stars in each field based on the Besan\c{c}on Galactic population model \citep[$N_{\rm BGM}$;][]{robin2003}. 
The Besan\c{c}on model counts are computed\footnote{http://model.obs-besancon.fr/} for the same magnitude and color range as the 
spectroscopic targets, and standard model parameters were used (models were drawn from
a 1 deg$^2$ line-of-sight centered on each mask, including all ages and spectral types).  

The scaled RGB counts are also corrected for two observational sources of bias: 
pre-selection of likely M31 RGB candidates using $M$, $T_2$, and 
DDO51 photometry ($c_{\rm DDO}$;
Equation (\ref{eqn:ddo}), Section~\ref{sec:filler}), and 
unequal sampling of the RGB luminosity function ($c_{\rm LF}$; 
Equation (\ref{eqn:imag}), Section~\ref{sec:lumfun}).  
Finally, the scaled and corrected RGB counts are 
converted to an $I$-band surface brightness.  The normalization 
is determined by fitting for an overall normalization factor, $I_{\rm norm}$, 
that best matches our surface brightness estimates to the minor axis 
$I$-band surface brightness profile 
published by \citet{courteau2011} (discussed further 
in Section~\ref{sec:fits}).  The normalization of the \citeauthor{courteau2011} profile is based 
on an $I$-band M31 image obtained by \citet{choi2002}. 
The surface brightness estimates are thus calculated as 
\begin{equation}
\Sigma_{\rm I} = \Sigma_{\rm norm} - 2.5\,{\rm log}_{10}\, \left[\frac{N_{\rm BGM}}{c_{\rm LF}\,c_{\rm DDO}}\,\frac{N_{\rm M31}}{N_{\rm MW}}\right].
\label{eqn:sb}
\end{equation} 

Our method of estimating the surface brightness in each field is based 
on counting the relative numbers of M31 RGB and MW dwarf stars.  
Selecting only securely identified M31 RGB and MW dwarf stars 
(Section~\ref{sec:cleansample}) 
is one way of doing this counting.  We 
have also tested including the marginal M31 RGB and MW dwarf candidates, and have investigated 
fitting the M31 RGB and MW dwarf star likelihood distributions (Figure~\ref{fig:likelihoods}) 
to find the fraction of stars in each population.  These alternate counting
methods produce similar results.  In the end however, our choice was influenced by the fact
that restricting the sample to stars securely identified as 
belonging to one population or the other provides a clean sample from which to fit for
kinematically cold components in each field. 

\subsection{Multiple Stellar Populations}\label{sec:sb_est_kcc}

\begin{figure}[tb!]
\plotone{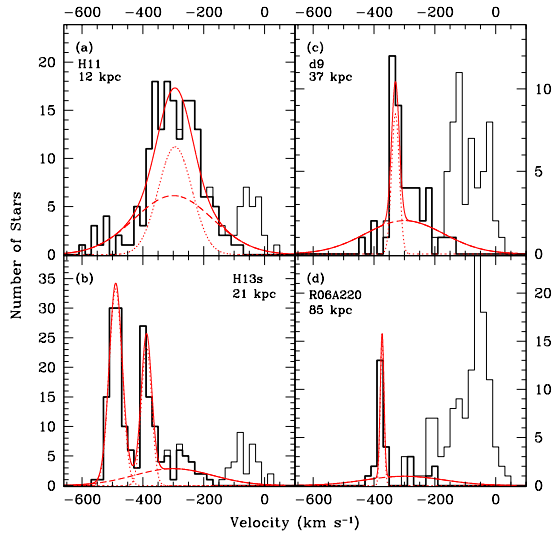}
\caption{Examples of multi-Gaussian fits (solid curves) to the velocity distribution 
of M31 stars (bold histogram) in fields with kinematically cold tidal debris.  
These fits are used to determine the fraction of stars belonging to 
M31's kinematically hot spheroid (Section~\ref{sec:sb_est_kcc}).  Thin histograms denote 
the velocity distribution of all stars with recovered velocities.  The fields shown 
span a range of projected distances from M31's center and demonstrate the 
range of velocity dispersions measured for the kinematically cold components 
detected in individual fields (dotted curves), from $\sim 5$~\kms\ (e.g., panel (d)) to $\sim 50$~\kms\
(panel (a)).  In some fields, more than one kinematically cold component 
is identified (panel (c)).  In all fits, the mean velocity and dispersion of the 
Gaussian representing M31's 
kinematically hot spheroid (dashed curves) is held fixed with parameters 
\mvsph$=-300$~\kms\ and \sigvsph$=129$~\kms\ \citep{gilbert2007}.  
}
\label{fig:mlgf_example}
\end{figure}

Many of the M31 halo fields targeted during the course of our
Keck/DEIMOS survey show evidence of multiple kinematical components 
within the M31 halo population: i.e., one or more distinct, kinematically cold
components associated with tidal debris in addition to M31's kinematically
broad spheroid \citep{guhathakurta2006,kalirai2006gss,gilbert2007,gilbert2009gss}.  
The effect of these fields on the surface brightness profile of M31's stellar 
halo will be discussed in detail below.

In fields with kinematically cold components 
(Table~\ref{tab:fields}), the number of M31 stars in the dynamically hot component
is calculated by multiplying the total number of M31 RGB stars in each field 
by the fraction of stars estimated to belong to the 
kinematically hot component.  This fraction is determined by calculating
maximum-likelihood multi-Gaussian fits to the stellar velocity distribution.  
The velocity distributions and associated 
maximum-likelihood fits of previously published fields with substructure are 
described in detail in the literature (Table~\ref{tab:fields}).

Figure~\ref{fig:mlgf_example} shows
examples of maximum-likelihood Gaussian fits in fields spanning a range 
of projected distances from M31's center.  
The analysis implicitly assumes the
presence of a dynamically hot component.  In all fields with identified kinematical
substructure, an underlying distribution of stars with a wide range of 
velocities is observed.   However, this does not preclude the possibility
that the broad distribution of stars (particularly in fields at larger radii) 
is a superposition of fainter, indistinct tidal debris features, rather than
a dynamically well-mixed population of stars. 

Only two of the new M31 halo fields have kinematical evidence of tidal debris.  
The velocity distribution of stars in these fields is presented in Figures~\ref{fig:mlgf_example}(c) and (d).  The substructure in field d9 is at \mvsb$=-329.7^{+10.3}_{-6.6}$~\kms, with \sigvsb=$12.6^{11.0}_{-4.8}$~\kms; $29.1^{+14.0}_{-12.6}$\% of  stars in this field are estimated to belong to the kinematically cold component.  
This field is located at 37~kpc near M31's major axis.  The expected disk velocity at this location is $\sim -100$~\kms\ \citep{ibata2005}, making it unlikely that this feature is related to M31's disk.  The peak of MW stars with $v\sim-130$~\kms\ may be related to the TriAnd overdensity in the MW halo; this feature is discussed further in \citet{tollerud2012}.  The substructure in field R06A220 is at \mvsb$=-373.5\pm3.0$~\kms, with \sigvsb=$6.1^{+2.7}_{-1.7}$~\kms; $42$\%$\pm15$\% of  stars in this field are estimated to belong to the kinematically cold component.   

Finally, we note that of the 39 fields presented here, only 4 specifically targeted tidal debris features: f207 and H13s targeted the giant southern stream, while the streamE and streamF fields were chosen to target photometric overdensities noted in \citet{tanaka2010}.  The spectroscopic data show clear kinematical substructure in f207 and H13s, but there is no detectable kinematical substructure in the streamE or streamF fields.    The remaining 35 fields were observed without prior knowledge of the presence of substructure in these fields, and thus represent a random sampling of the properties of M31's stellar halo.   

\subsection{Primary Sources of Bias and Systematic Error}\label{sec:biases}
Since the surface brightness estimates are based on counts of spectroscopically
confirmed samples of M31 RGB and MW dwarf stars, nonuniform spectroscopic 
target selection and variable observing conditions can significantly bias our
measurements. 
Radially dependent systematic errors that affect the slope of the measured
surface brightness profile are the most important to consider, since a 
purely empirical normalization is used to convert counts of confirmed
M31 RGB stars and MW dwarf stars to surface brightness estimates. 
The surface brightness estimates are corrected 
for two sources of bias: selection of likely M31 RGB stars for spectroscopy
based on Washington $M$, $T_2$, and DDO51 photometry, and the luminosity
function (LF) of spectroscopic targets with successful velocity measurements.  
We describe each of these in
detail below (Section~\ref{sec:filler} and {sec:lumfun}).   
Other sources of error that are not accounted for are discussed
in Section~\ref{sec:othererrors}. 

\subsubsection{DDO51 Selection Efficiency}\label{sec:filler}
Selection of likely M31 RGB stars using photometry in the 
Washington $M$ and $T_2$ filters and the DDO51 filter (Section~\ref{sec:obs})
boosts the measured RGB/dwarf star ratios relative to fields 
without DDO51 photometry, and results in an offset 
in the surface brightness between fields with and without DDO51 
photometry \citep{guhathakurta2005}. 
Among the masks that benefit from DDO51-based target selection, 
masks at large \rproj\ contain a higher fraction of ``filler'' targets 
(that fail the DDO51 criterion for RGB stars)
than masks at small \rproj.  Since many of these filler targets are MW
dwarfs, this effect reduces the measured M31 RGB/MW dwarf ratio.  
Uncorrected, this would cause the derived radial surface brightness profile 
to be steeper (a more negative power-law index) than the true profile
because a larger fraction of slits will target likely M31 RGB stars in DDO51-based 
masks at small \rproj. 

To correct for this effect, we estimate the factor by which the 
measured RGB/dwarf star ratios are increased by DDO51-based target selection. 
For each field a DDO51 selection function (DSF) is computed, defined as 
the ratio of the number of stars selected for inclusion on a 
spectroscopic mask ($N_{\rm target}$) to the number of stars available ($N_{\rm total}$), as a function
of the DDO51 parameter ($f_{\rm DDO51}$; Figure~\ref{fig:global_diags}(b)).  The $f_{\rm DDO51}$ value
measures the probability a star is an M31 red giant 
based on the Washington $M$, $T_2$, and DDO51 photometry \citep{palma2003,majewski2005}. 
The DDO51 correction factor ($c_{\rm DDO}$) is then defined as the ratio of two integrals:
(1) the M31 RGB DDO51 probability distribution function 
(${\rm PDF_{RGB}}$, Section~\ref{sec:cleansample}),  
convolved with the DDO51 selection function,
and (2) the MW dwarf star DDO51 PDF (${\rm PDF_{dwarf}}$) convolved with the DDO51 selection 
function. The DDO51 PDFs for M31 and MW stars are 
shown in Figure~\ref{fig:global_diags}(b). 
\begin{eqnarray}
{\rm DSF}(f_{\rm DDO51}) = \frac{N_{\rm target}(f_{\rm DDO51})}{N_{\rm total}(f_{\rm DDO51})} \\
c_{\rm DDO}  
= \frac{\int{\rm PDF_{RGB}} \, {\rm DSF} \, df_{\rm DDO51}}{\int{\rm PDF_{dwarf}} \, {\rm DSF} \, df_{\rm DDO51}}
\label{eqn:ddo}
\end{eqnarray}
The scaled M31 RGB counts in each field are 
divided by the DDO51 correction factor to account for differences in the DDO51 selection
function between fields (Equation \ref{eqn:sb}).  For fields in which DDO51 photometry was not used for 
designing the masks, $c_{\rm DDO}$ is set to unity.  

In field a0 (\rproj\,$\sim 30$~kpc), an inner field with many high priority 
targets, the DDO51-based selection of spectroscopic targets increased 
the measured RGB/dwarf ratio by a factor of 3.4.  In the outermost fields 
where there are only a few bright stars ($I_0<23.0$) 
with high DDO51 parameters per mask, the DDO51-based selection 
increased the measured RGB/dwarf star ratio by a factor of 1.4.  

The application of the DDO51 correction factor to the M31 RGB/MW dwarf ratio 
results in excellent agreement between the surface brightness 
estimates of fields with and without DDO51-based target selection at 
\rproj\,$\sim 30$~kpc (e.g., `mask4' (no DDO51) and `a0' and `a3' (DDO51)).  
Interior to this radius fields 
did not have DDO51 photometry, while exterior to this radius the 
majority of the fields do have DDO51 photometry.  The DDO51 correction factor 
removes the need to empirically adjust the normalizations of the surface 
brightness estimates for fields with and without DDO51 photometry. 
   
\subsubsection{Sampling of the Stellar Luminosity Function}\label{sec:lumfun}

In designing the spectroscopic masks, higher priority is given to 
brighter targets because recovering a velocity for 
these objects is more likely.  However, a mask at large \rproj\ typically contains a 
higher fraction of faint targets than a mask at small \rproj\ due to the 
sparseness of stars in M31's outer halo.  For example, the
number of $20.5<I_0<22$ targets is about twice that of $22<I_0<22.5$ 
targets in the \rproj\,$=30$ and 60~kpc fields a0 and a13, while these 
two magnitude ranges contain comparable numbers of targets in the outer 
fields m8 and m11 (at 120 and 165~kpc).  

The M31 RGB $I$-band LF rises more steeply toward faint
magnitudes than the MW dwarf $I$-band LF.  Therefore the measured 
RGB/dwarf star ratio is expected to be higher in fields with a larger number 
of faint targets.  If uncorrected, the larger number 
of faint targets in the outermost fields would bias the measured surface 
brightness profile to be flatter than the true profile.  The amount of bias is limited because 
the spectroscopic targets span a relatively small apparent magnitude range and
the radial velocity measurement fails for many stars with $I_0>22$, due to
low S/N in the continuum.

In addition, variations in observing conditions
(e.g., seeing and transparency) from one mask to another can lead to variations
in the completeness function of radial velocity measurements.  We attempted
to minimize this effect by increasing the total exposure time for masks
observed under sub-optimal conditions.  Nevertheless there are
field-to-field variations in the faint-end limiting magnitude of the
spectra, which in turn leads to slight
variations in the fraction of the LF that is included in our
statistics.

To account for the above effects, we estimate the factor by which
the measured M31 RGB/MW dwarf star ratio is increased in each field due to 
(1) a larger fraction of faint targets and/or (2) a fainter limiting magnitude 
for recovery of radial velocities from the spectra.  
The $I$ magnitude 
recovery function (IRF) is defined as the ratio of the number of stars with
successful radial velocity measurements ($N_{\rm vel}$) to the total number of 
available targets in the photometric catalog ($N_{\rm total}$), as a function of $I$ 
magnitude.  The correction factor ($c_{\rm LF}$) for each field 
is defined as the ratio of two integrals: (1) the M31 RGB LF convolved with the $I$ magnitude recovery function, and 
(2) the MW dwarf star LF convolved with the $I$ magnitude recovery function.
\begin{eqnarray}
{\rm IRF}(I) = \frac{N_{\rm vel}(I)}{N_{\rm total}(I)} \\
c_{\rm LF} = \frac{\int {\rm LF_{M31}}\, {\rm IRF}\, dI}{\int {\rm LF_{MW}}\, {\rm IRF}\, dI}
\label{eqn:imag}
\end{eqnarray}

This is analogous to the procedure used for calculating the DDO51 selection 
efficiency factor (Section~\ref{sec:filler}). 
The LF of M31 RGB stars is assumed to be $dN/dI=10^{0.3I_0}$, with a 
cutoff at $I_0=20.5$ corresponding to the
tip of the RGB.  The MW dwarf star LF is assumed to be a 
constantly rising function over the magnitude range of the spectroscopic 
survey; the slope is based on the Besan\c{c}on Galactic population model 
\citep{robin2003}. 

The scaled M31 RGB counts in each field are divided by
the correction factor $c_{\rm LF}$ to account for differences in 
the target selection and limiting magnitude of the spectroscopic 
observations in each field.  The correction for sampling of the LF
is a smaller correction than the correction for DDO51 selection:
while the maximum variation in the DDO51 correction is a factor of $\sim 4$, the  
maximum variation in the LF correction is a factor of $\sim 2$.

\subsection{Other Sources of Error}\label{sec:othererrors}
There are additional potential sources of error in the surface brightness
estimates.  We briefly discuss these and their possible impacts on
the results below.  
In general, their effect on the measured surface brightness profile 
will be to add an extra source of noise in the measurements.  

\subsubsection{Masks Targeting dSphs}
Some of the M31 fields target dSph galaxies in M31's halo 
(Section~\ref{sec:dsph_fields}).  
In the dSph masks, the precision with which the number of field stars in 
M31's halo can be estimated depends on both the spatial extent of the 
dSph and the tightness of the locus of dSph member stars in the 
$v_{\rm los}$\,--\,[Fe/H] plane.  The amount of parameter
space occupied by dSph members limits the parameter space where
M31 field stars can be securely identified.  In most cases, the parameter space
spanned by the dSph is small compared to the parameter space spanned by M31 halo stars. 
The magnitude of the error in M31 RGB star counts thus introduced can be 
estimated by counting the number of M31 stars in the range of [Fe/H] values and  
width in velocity occupied by dSph members, but on the opposite side of M31's 
Gaussian velocity distribution (i.e., if the dSph has a mean velocity of $-450$~\kms, 
one can count the number of M31 stars at $-150$~\kms\ that would have been within 
the area of the $v_{\rm los}$\,--\,[Fe/H] locus used to identify dSph stars).  
This additional source of error is generally found to be significantly smaller than the Poisson error; at the most, it is comparable in fields with very few M31 RGB stars.

\subsubsection{Population Gradients}
The PDFs used to identify M31 RGB stars were defined using
a training set that was primarily drawn
from fields in the inner, metal-rich regions of M31's stellar halo.  
If the RGB classification method is significantly biased against selecting metal-poor stars, 
it is possible that large scale radial gradients in the metallicity of the 
halo population might bias the profile measurement.  M31's stellar halo does
grow increasingly metal-poor with increasing radius \citep{kalirai2006halo, koch2008,tanaka2010}.  A gradient in the age of 
the stellar population would also produce a bias, as changes in 
age produce similar color offsets as changes in metallicity for stars on the 
RGB.  

However, there does not appear to be a significant bias
against identifying metal-poor RGB stars as M31 stars in the sample.  
If the color-based and metallicity-based diagnostics biased the selection against blue,
metal-poor stars, it would be most apparent in the velocity range $-200$ to $-150$~\kms.  
This is where
the PDFs of the M31 RGB and MW dwarf velocity diagnostics 
overlap (Figure~\ref{fig:global_diags}) and 
the effect of the velocity diagnostic on the overall likelihood is minimized.  
Figure~\ref{fig:likelihoods} shows that the range of velocities spanned by stars 
classified as secure M31 red giants is similar whether the stars have \feh$>-1$ or \feh$<-1$.

\subsubsection{Photometric Data Quality}
Field-to-field variations in 
photometric accuracy and image quality lead to slight
variations in the selection efficiency of M31 RGB spectroscopic candidates 
and the fraction of contaminating foreground MW dwarf stars and 
compact background galaxies observed on the spectroscopic masks.  
In general, the limiting magnitude of the photometric data is significantly fainter 
than that of the spectroscopic data, thus the error introduced by this effect is 
expected to be small compared to the Poisson errors.  

\subsubsection{Galactic Star-count Model}
The surface density of MW dwarf
stars predicted by the Besan\c{c}on Galactic population model \citep{robin2003} only
enters the final surface brightness estimates in a relative sense.  It is used solely as
a normalization factor ($N_{\rm BGM}$) to account for the changing density of MW dwarf stars across
the survey, and a separate normalization factor is used to convert the surface densities to
surface brightnesses (Section~\ref{sec:sbprofile}).  Therefore, an absolute error in the total number 
of stars predicted by the Besan\c{c}on model would not affect the resulting surface brightness 
estimates.

However, the Besan\c{c}on model provides only a first order approximation for the true
changes in MW stellar density across the footprint of the survey.  A
field-dependent error in the predicted projected density of MW dwarf stars
could affect the resulting surface brightness profile.   

If the Galactic latitude dependence of the number counts in the model is incorrect, 
this would result in a systematic error in the surface brightness estimates with a 
magnitude that varies as a function of the Galactic latitude of the field.  A factor of two 
systematic error in the Besan\c{c}on model between the highest
and lowest Galactic latitude fields would result in a maximum systematic error of 
0.75 mag; a factor of 1.5 error in the model counts would result in a maximum systematic 
error of 0.4 mag, which is comparable to the  
Poisson uncertainties for many of the fields.  

However, there is not a direct correspondence
between the projected distance of the fields from M31's center and their Galactic latitude.  Thus, the
primary effect of a systematic error in the Galactic latitude dependence of the Besan\c{c}on model number counts will be increased scatter in the surface brightness profile rather than a bias 
in the slope of the profile fit.

The Besan\c{c}on model accounts only for the smooth components of the MW.  However, the MW's halo is known to have abundant substructure, and there is one known substructure, the Triangulum-Andromeda feature \citep{rocha-pinto2004}, which is in the direction of M31 and may affect some of the fields.  Since the increased MW stellar density due to substructure is not accounted for by $N_{\rm BGM}$, the subsequent increase in observed MW star counts would result in a fainter M31 halo surface brightness estimate for a line of sight that passes through MW substructure.  Therefore, the net effect of significant MW halo substructure will be increased scatter in the observed M31 halo surface brightness profile.

\section{Surface Brightness Profile of M31's Stellar Halo}\label{sec:sbprofile}
\begin{figure*}[tb]
\centerline{
\includegraphics[width=0.5\textwidth]{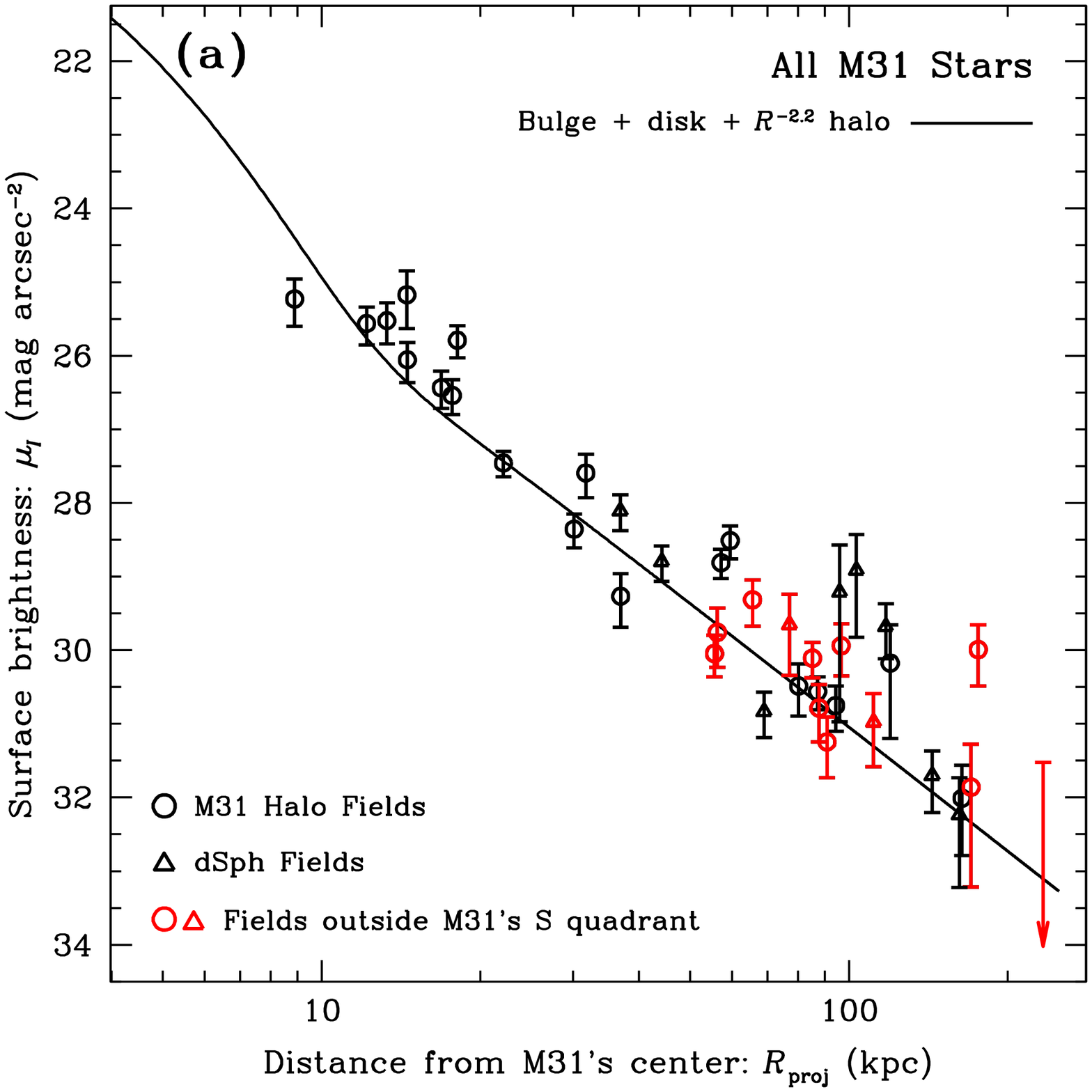}
\includegraphics[width=0.5\textwidth]{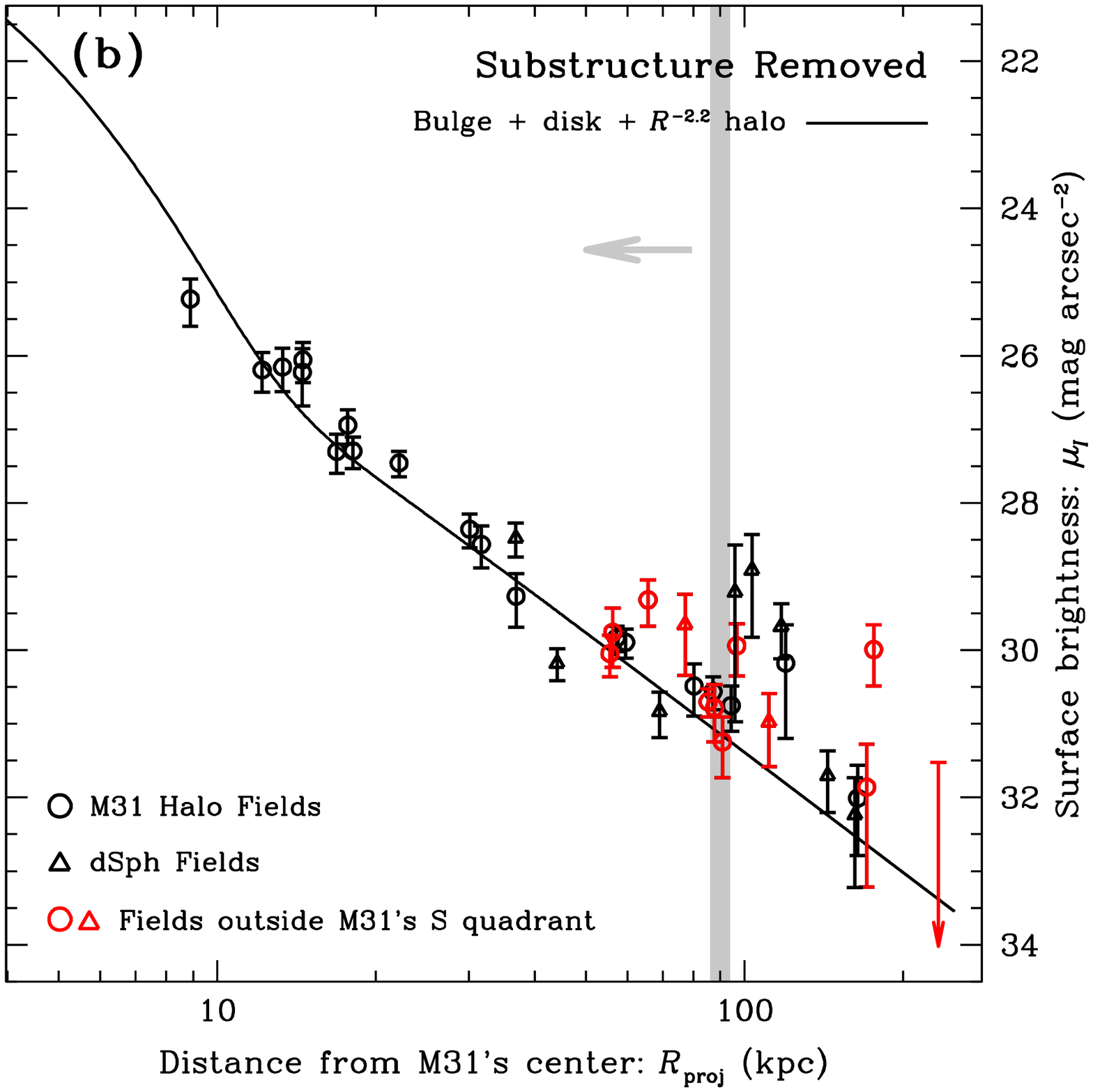}
}
\caption{Surface brightness profile of M31's halo 
(Section~\ref{sec:sbprofile}), based 
on counts of spectroscopically confirmed RGB stars (Section~\ref{sec:sb_est}).  
(a) Surface brightness estimates based on all M31 RGB stars in a 
given field.  
(b) Surface brightness estimates of the 
kinematically hot population of M31 RGB stars in each field  (Section~\ref{sec:sb_est_kcc}); due to 
the sparse nature of the outermost regions,  
tidal debris can only be identified in fields within \rproj\,$\lesssim 90$~kpc.  
The statistical subtraction of identified tidal 
debris features results in a smoother surface brightness 
profile for M31's stellar halo.  
Error bars are the combined Poisson 
errors of the M31 RGB and MW dwarf star counts.  For fields with substructure statistically
removed, the error in the M31 RGB counts includes the uncertainty in the 
fraction of the population belonging to the kinematically hot component.    
The solid curve shows the maximum-likelihood minor axis surface brightness 
profile fit to the data and includes a S\'ersic bulge, 
exponential disk, and power-law halo.  The bulge and disk component 
parameters are set to those found to best fit a compilation of 
observations of M31's light 
distribution by \citet{courteau2011} (Section~\ref{sec:fits}). 
Red points denote measurements from fields outside the southern
quadrant of M31's halo, confirming that M31's stellar halo 
extends to large projected distances (\rproj\,$>150$~kpc) and follows the same 
power-law decline in surface brightness in all directions 
(Section~\ref{sec:nquad}).   
}
\label{fig:sb}
\end{figure*}

Figure~\ref{fig:sb} displays the surface brightness estimates (Table~\ref{tab:sbest}) as a 
function of projected distance from M31's center for each of the M31 
fields shown in Figure~\ref{fig:roadmap}.  
The surface brightness profile of M31's stellar halo clearly extends to large 
projected distances (\rproj\,$\gtrsim 175$~kpc), with no evidence of a downward 
break in the profile 
(i.e., a transition to a steeper profile) at large radii.

To take into account the flattened 
profile of M31's inner regions, the \rproj\ values of a small subset of the fields 
were converted to an effective radial distance along M31's minor axis. 
This effective \rproj\ was calculated adopting a flattening of $b/a = 0.6$ 
for \rproj\,$< 30$ kpc, as measured for M31's inner spheroid 
\citep{walterbos1988,pritchet1994}. 
No correction was necessary for the majority of the innermost fields 
(\rproj\,$\lesssim 30$~kpc), since most lie along M31's minor axis.
There are no strong observational constraints on the ellipticity 
of M31's spheroid at large distances, therefore the 
spheroid of M31 is assumed to be spherical ($b/a = 1$) 
beyond \rproj\,$=30$~kpc.  This assumption is tested in Section~\ref{sec:shape}.

Approximately a third of the fields targeted 
during the course of the SPLASH Keck/DEIMOS survey have conclusive kinematical
evidence of tidal debris \citep[Table~\ref{tab:fields};][]{guhathakurta2006,kalirai2006gss,gilbert2007,gilbert2009gss}.  
In many cases, these kinematically cold components can be associated 
with known tidal debris features identified as photometric overdensities in 
star count maps.  The left panel of Figure~\ref{fig:sb} shows the surface
brightness estimates including all M31 halo stars, while the right panel
shows the surface brightness estimates with the kinematically cold tidal
debris features statistically subtracted (Section~\ref{sec:sb_est_kcc}).  In fields with 
kinematically cold tidal debris features, the error bars shown in Figure~\ref{fig:sb}(b)
include the uncertainty
in the number of RGB stars belonging to the kinematically hot component, estimated 
from the maximum-likelihood fits to the field's stellar velocity 
distribution (Section~\ref{sec:sb_est}).  
 
Statistical subtraction of tidal debris features 
reduces the scatter about the best-fit profile for the inner 
and medial fields; this effect is quantified below (Section~\ref{sec:fits}). 
Significant scatter in the surface brightness estimates is 
noticeable at \rproj\,$\sim 90$~kpc (Figure~\ref{fig:sb}), 
particularly in the right panel where tidal debris features
have been statistically subtracted from the M31 RGB counts.  
The most distant field in which  
kinematically cold components can be identified in the spectroscopic data is at 
\rproj\,$= 85$~kpc (Table~\ref{tab:fields}).  In fields at these large projected
distances, there are typically $\lesssim 10$ secure M31 RGB stars per 
field.  This makes discovering a tidal debris feature in the spectroscopic 
data prohibitively difficult unless it happens to dominate the line of sight.  
It is likely that there is substructure in some of these outer fields 
that cannot be 
identified as such due to the limited number of M31 stars.  For 
example, fields `streamE' and `streamF' were placed on photometric 
overdensities identified as tidal debris features by \citet{tanaka2010}. 
However there is no definitive evidence of kinematically cold components in the 
spectroscopic data.

In the following subsections, we will discuss fits to the surface 
brightness profile of M31 with and without the inclusion of 
kinematically-identified tidal debris (Section~\ref{sec:fits})
and explore the extent of M31's stellar halo (Section~\ref{sec:extent}) 
including an examination of the M31 RGB sample for signs of MW dwarf 
star contamination (Section~\ref{sec:MWcontamination}).
We also discuss the scatter of the data about the
profile fit (Section~\ref{sec:scatterwithsubstructure}), measure the ellipticity of M31's
stellar halo (Section~\ref{sec:shape}),
and compare our
surface brightness estimates with M31 halo profile fits from the 
literature (Section~\ref{sec:compare}).

\subsection{Fits to Profiles With and Without Tidal Debris Features}\label{sec:fits}

The innermost data points (\rproj$\lesssim 20$~kpc) are well within the region of M31's halo where previous studies have found the profile to be consistent with an extension of M31's bulge (Section~\ref{sec:intro}).
Therefore, we do not assume that contributions to the surface brightness from 
M31's bulge and disk are non-negligible in these fields.
A combined bulge, disk and power-law halo profile is fit
to the M31 surface brightness estimates (Section~\ref{sec:sb_est})
using a maximum-likelihood analysis.  The intensity at a given radius is
given by the functional form
\begin{equation}
I=I_{\rm b}(R_{\rm proj}) + I_{\rm d}(R_{\rm proj}) + I_0\left(\frac{1+(R_*/a_h)^2}{1+(R_{\rm proj}/a_h)^2}\right)^\alpha
\label{eqn:profile}
\end{equation}
where $I_b$ and $I_d$ represent the profile of M31's bulge and disk  
along the minor axis,
$-2\alpha$ is the halo's two-dimensional power-law index,  $a_h$ is a core radius inside of which the halo 
profile flattens, and $I_0$ and $R_*$ are normalization constants.
  
The recently published M31 surface brightness profile 
by \citet{courteau2011} is used to fix the profiles of M31's bulge ($I_{\rm b}$) and disk ($I_{\rm d}$). 
\citeauthor{courteau2011}\ 
analyzed photometric observations of M31's inner regions, and compiled them
with previously published surface brightness estimates to measure M31's surface 
brightness profile along the major and minor axes.  Here, we use their preferred
minor-axis fit ``U,'' which included a S\'ersic bulge ($n=1.9$, $R_e=0.73$~kpc), 
exponential disk (with scale length $R_{\rm d}=5.0$~kpc), 
and power-law halo.  While this particular bulge/disk decomposition reproduces
the observed surface brightness profile along the minor axis, the 
kinematical data in the innermost fields do not support the presence of such a prominent disk 
component \citep{gilbert2007}; an attempt to include the kinematical data as a constraint 
on the inner minor-axis profile will be the subject of a future paper (Dorman et al., in preparation). 

Although this work has significantly more data at large radii 
than \citeauthor{courteau2011}, their minor axis profile 
has a much higher density of points out to 30~kpc. 
Their profile thus provides significantly more leverage
for fitting the transition from M31's inner regions (bulge and disk) to 
M31's halo than the data presented here.  Therefore, we adopt the 
core radius of $a_h=5.2$~kpc found for the \citeauthor{courteau2011}\ `U' fit. 
This parameter sets the amplitude of the halo
at small radii, and affects the radius at which the halo profile turns over.  
We adopt $R_*=30$~kpc, which is safely in the regime where the halo 
dominates the surface brightness profile; this choice is arbitrary and 
only affects the value of $I_0$ ($I_h(R_*)=I_0$).

The total model intensity is compared to the surface brightness estimates 
to determine 
the power-law parameters (normalization, $I_0$, and power-law index, $-2\alpha$)
that provide the maximum-likelihood fit to the data.   
The errors in the surface brightness
estimates are used as weights when performing the maximum-likelihood analysis. 
When calculating the fit to the data set with kinematically identified tidal debris 
statistically removed (Figure~\ref{fig:sb}(b)), the fit was restricted to the 
radial region over which each field has sufficient numbers of M31 stars to identify 
kinematical substructure (\rproj\,$<90$~kpc).  

Given the significant scatter in the data, especially at large radius, 
we explore the effect of individual surface brightness estimates and 
their errors 
on the best-fit power-law index using Monte Carlo tests.  Each trial was 
randomly drawn from the observed sample, allowing 
redrawing of individual data points.  For each data point 
drawn, the number of M31 RGB and MW dwarf stars
used to determine the surface brightness estimate are adjusted to
a random deviate drawn from a Poisson distribution.
The surface brightness estimates are then recalculated, and the 
maximum-likelihood power-law profile determined; 1000 trials were performed.

The data are best fit by a power-law profile with index $-2.2\pm0.2$ and
normalization $\mu_0=-2.5\,{\rm log}\,I_0=28.15\pm0.15$~mag arcsec$^{-2}$. 
The quoted error estimate is the weighted standard deviation of the distribution of 
best-fit power-law indices from the Monte Carlo tests. 
This is formally consistent with the best-fit power-law found by 
\citeauthor{courteau2011}\ (index $-2.5\pm0.2$ and $\mu_0=28.07\pm0.05$~mag arcsec$^{-2}$). 
The best fit power-law 
profile index for the sample with kinematically cold tidal debris 
features removed is also {\bf $-2.2\pm0.3$}, fully consistent with the index found 
using all stars, but with a fainter halo normalization: $\mu_0=28.58\pm0.14$~mag arcsec$^{-2}$.  

The measured halo profiles are not very sensitive to the adopted
bulge and disk profiles.  Even if the surface
brightness is assumed to be dominated by the halo for the full range of the data, 
and only a power-law component is fit, the best-fit index is $-2.4\pm0.2$, 
consistent with the power-law fit including disk and bulge
components.  The power-law index is also fairly insensitive to the adopted
core radius.  If a smaller core radius, $a_h=2.7$~kpc, is adopted, 
consistent with an M31 halo measurement
based on blue horizontal branch stars as tracers of the halo \citep{williams2012}, 
the best-fit index is $-2.3\pm0.2$. 

The best-fit index is also not strongly
dependent on the radial range used for fitting the data.  If    
the radial range is restricted to \rproj\,$>20$~kpc, the best-fit 
power-law index is $-2.1\pm0.3$ for the sample including all M31 stars;
the best-fit power-law index is $-1.9\pm0.4$ for \rproj\,$>35$~kpc.  
Although these best-fit values are shallower than the best-fit profile to all data points,
the power-law indices are all consistent within the errors.
The best-fit index for the sample with kinematical substructure statistically removed in the  
radial range $20<$\,\rproj\,$<90$~kpc is also consistent: $-2.0\pm0.5$.

\subsection{Extent of M31's Halo}\label{sec:extent}

\subsubsection{Constraints on Contamination from MW Dwarf Stars}\label{sec:MWcontamination}
\begin{figure*}[tb!]
\centerline{
\includegraphics[width=0.5\textwidth]{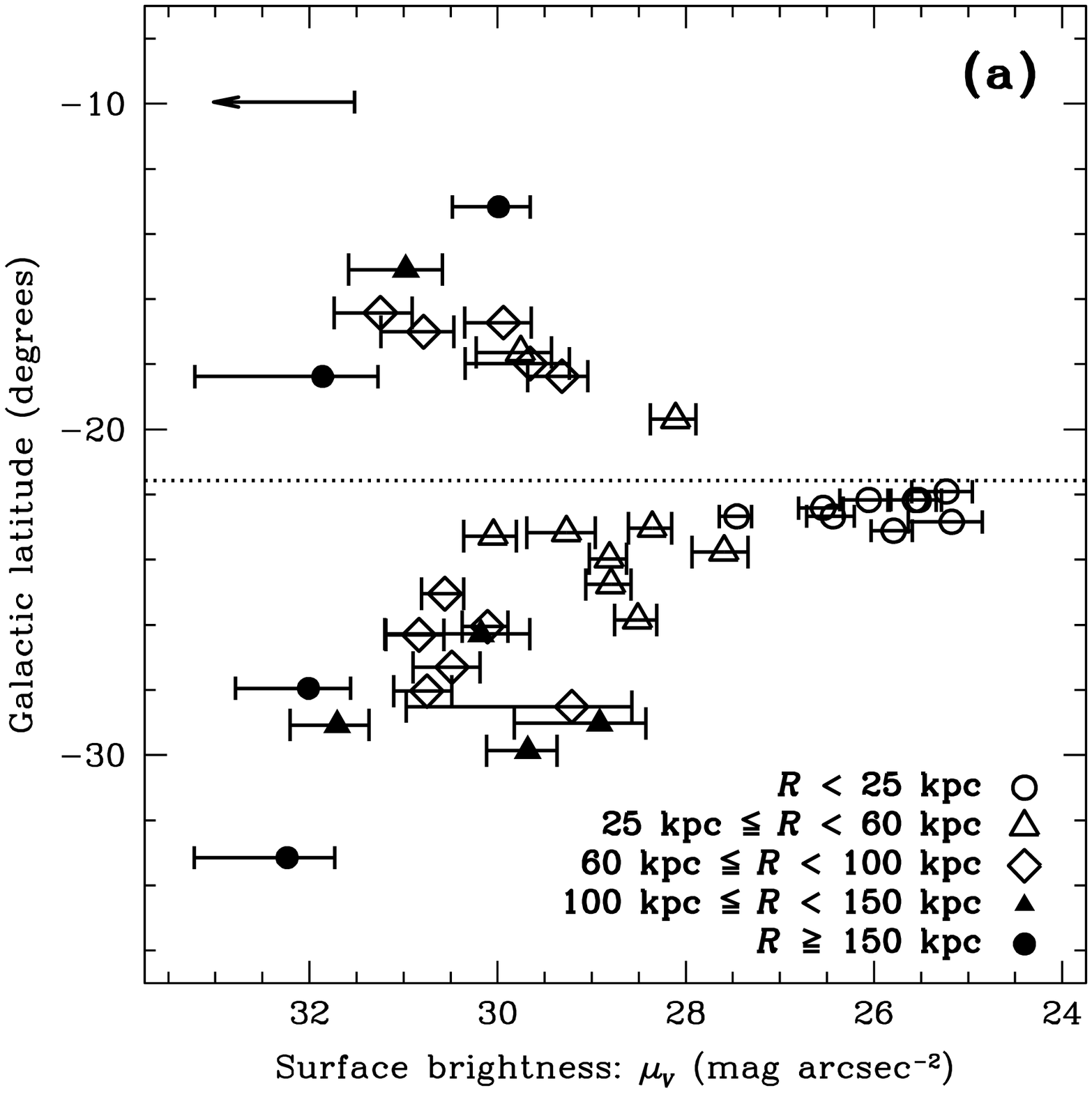}
\includegraphics[width=0.5\textwidth]{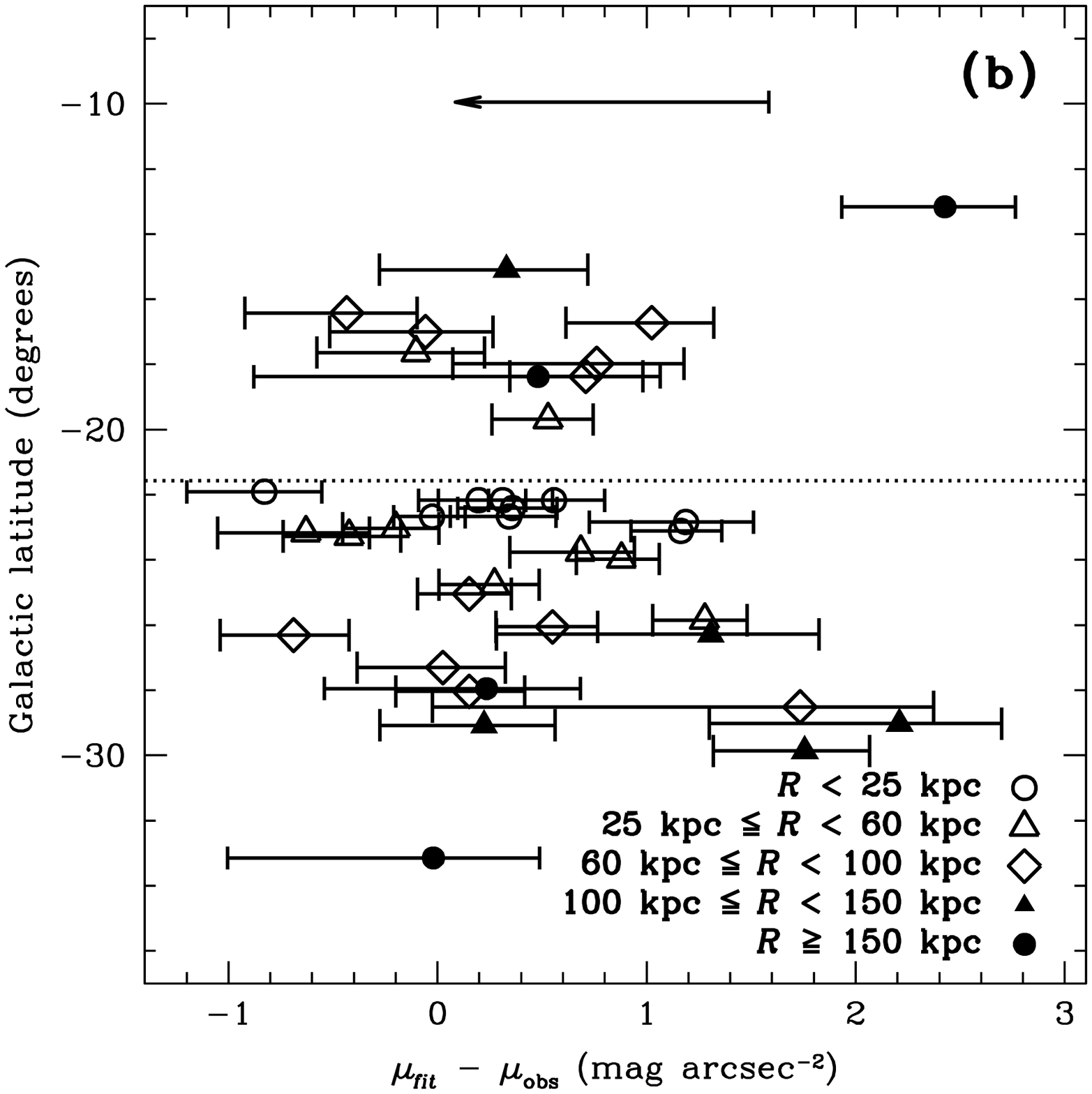}
}
\caption{Surface brightness of the M31 spectroscopic fields as a 
function of Galactic latitude.  The left panel (a) shows measured surface brightness
estimates, while the right panel (b) shows the scatter of the surface brightness estimates 
about the profile fit.  The symbols used vary according to the 
radial distance of the field from M31's center.  The dotted line denotes the 
Galactic Latitude of M31's center.  If contamination from foreground MW disk stars 
was a significant factor in the M31 RGB surface brightness
measurements, fields at lower absolute
galactic latitude (closer to the disk of the MW) would be expected to have 
systematically higher surface brightness estimates than fields at 
comparable radii but higher absolute galactic latitude.  
Significant contamination would be most noticeable in the 
fields at large projected distances in M31's halo, where the number of detected M31 RGB 
stars is small.
No trend is seen in the data, indicating that foreground MW disk
stars are not a significant source of contamination in the surface 
brightness estimates of the outer halo fields (Section~\ref{sec:MWcontamination}).    
}
\label{fig:mu_gallat}
\end{figure*}

M31 RGB stars are detected as far as $175$~kpc from the center of 
M31.  The most distant field included in this paper, d7 (\rproj$ = 230$~kpc), 
has objects identified as marginal M31 stars, but no secure M31 RGB star 
detections (Section~\ref{sec:cleansample}).  
Given the large projected radial distances of the outermost fields, it 
is prudent to explore the effects that significant MW dwarf star 
contamination would have on the surface brightness estimates.
Foreground MW dwarf star contamination consists of two physical groups of 
stars: those belonging to the MW's disk and halo.  

Dwarf stars in the disk of the MW have 
heliocentric line-of-sight velocities near $\sim 0$~\kms\ (Figure~\ref{fig:likelihoods}).
The surface density of MW disk stars increases as the galactic latitude of the field approaches 
0\degree; this effect is significant over the area spanned by our spectroscopic survey (Figure~\ref{fig:roadmap}).  If MW disk stars are a significant source of contamination in the M31 RGB 
sample, fields closer to the MW disk (at lower absolute galactic latitude) should have 
systematically higher surface brightness estimates than fields farther removed from the MW disk. 
No clear trend is seen in the data (Figure~\ref{fig:mu_gallat}, left panel), although each radial bin encompasses a rather large range of radii and hence intrinsic surface brightnesses.  

The effect of the variation of surface brightness as a 
function of radius can be accounted for by subtracting the measured $\mu_{I}$ from the 
best fit profile (Figure~\ref{fig:mu_gallat}, right panel).  Intrinsic 
scatter of the M31 halo's surface brightness about the profile fit will 
be independent of galactic latitude, producing a pure scatter plot.   
If significant contamination from MW disk stars was present in the sample, 
there would be a clear trend in $\mu_{\rm fit} - \mu_{\rm obs}$ as a function
of Galactic latitude, with 
preferentially larger values at lower absolute Galactic latitudes.  
Based on the Besan\c{c}on Galactic model \citep{robin2003}, the density 
of MW stars along the line of sight is expected to increase by a factor 
of $\sim 8$ between the lowest and highest galactic latitude fields.  
If an approximately constant fraction of the MW disk stars 
present in each field creeps into the M31 RGB sample, the 
expected difference in surface brightness would be 2.2 magnitudes.  
Restricting the range of galactic latitudes to where the majority of the 
data lie ($-30<b<-15$) leads to an expected difference of 1.3 magnitudes.  
There is no apparent trend in the data even in the two outermost radial 
bins, let alone a trend this large.  Therefore, we conclude that 
MW disk star contamination is not significantly affecting the M31 RGB 
surface brightness estimates.  This also further demonstrates that the 
method of separating M31 RGB and MW dwarf stars (Section~\ref{sec:cleansample}) 
is highly effective.      

Bright, distant main sequence turn-off stars in the MW's halo have a 
large range of velocities, and they can have very negative 
(M31-like) line-of-sight velocities (Figure~\ref{fig:likelihoods}).  
Their relatively blue colors make them particularly difficult to 
distinguish from M31 RGB stars, since the diagnostics described 
in Section~\ref{sec:cleansample}
have less power at bluer colors (Figure~\ref{fig:global_diags}).
Foreground MW halo stars are expected to have approximately constant surface
density across the area spanned by our M31 spectroscopic survey.  Significant
contamination from foreground MW halo stars in the M31 RGB halo sample 
would therefore manifest itself as a relatively constant (additive) surface 
brightness term, irrespective of radius. Naively, a constant term 
would simply be adjusted for in the empirical normalization of the data.  
However, given the spectroscopic selection function, we
observe more MW halo stars on masks at large radii (because more 
filler targets are placed on the masks), and contamination is therefore
expected to be larger in the outermost regions.  If the term were large, 
it would therefore decrease the slope of the observed surface brightness 
profile. 

At some large radius, this term would become larger than
the signal of actual M31 RGB stars, and the measurements would become 
foreground-limited, which would manifest as a constant observed surface 
brightness with increasing radius.  However, the surface brightness estimates 
continue to decrease with radius, consistent with the power-law profile (although 
with increasing scatter), out to the largest radii probed 
in our survey (Figure~\ref{fig:sb}).   
Furthermore, the M31 halo sample and the likely MW halo stars (stars significantly bluer than
the most metal-poor RGB isochrone, Figure~\ref{fig:likelihoods}) occupy distinct
regions of the \feh\,--\,$v_{\rm los}$ plane.

\begin{figure}[tb!]
\plotone{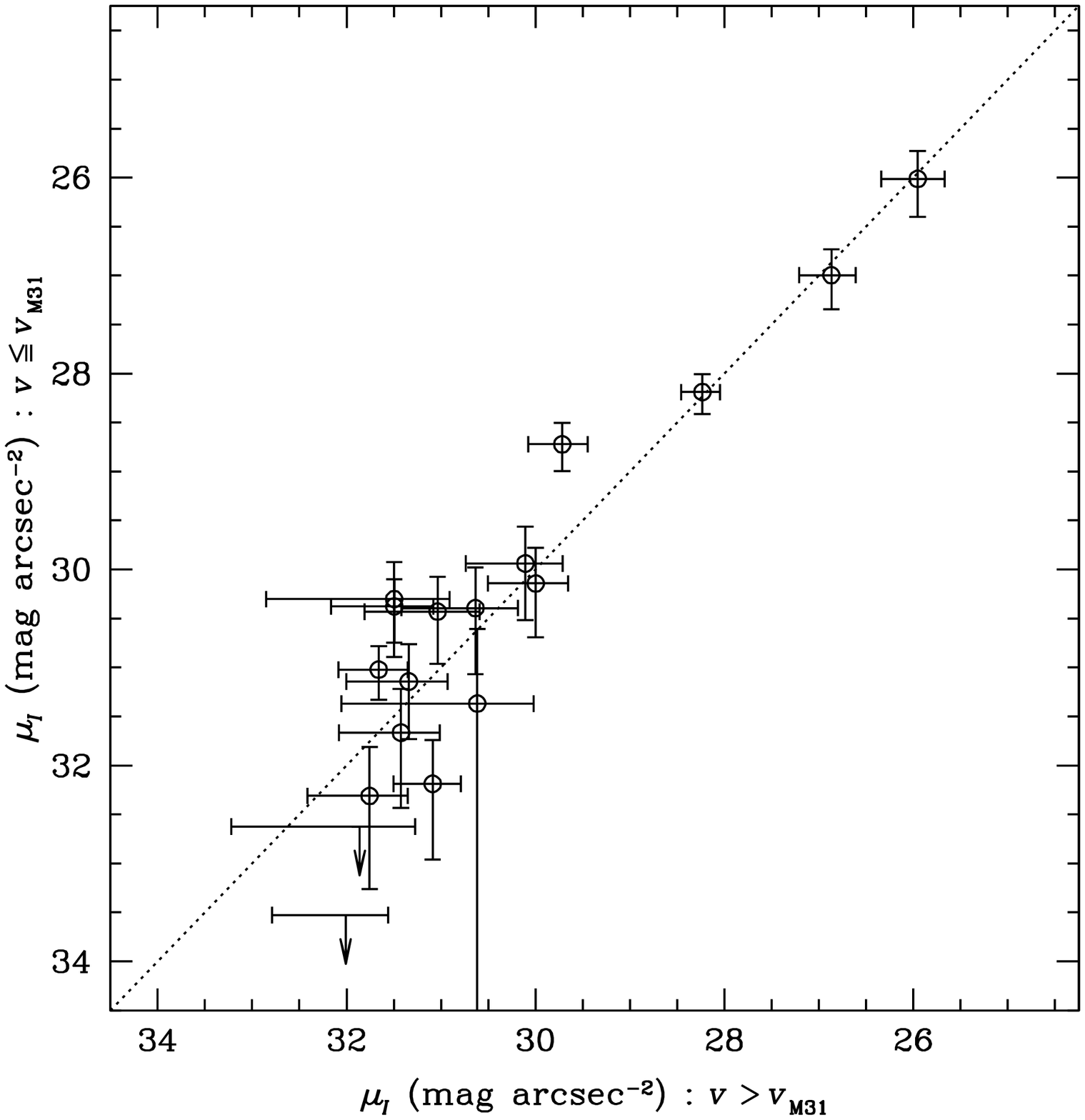}
\caption{Surface brightness estimates of a subset of the M31 spectroscopic fields, calculated after splitting the M31 RGB sample at the systemic velocity of M31 ($v_{\rm M31}=-300$~\kms).  Only fields with a reasonable expectation of symmetry in the M31 velocity distribution are shown here; this excludes fields with kinematically identified tidal debris features and containing dwarf spheroidal galaxies (Table~\ref{tab:fields}).    The dotted line denotes the one-to-one line.  MW halo stars have a broad range of velocities centered at $v\sim-175$~\kms.  Thus, significant contamination from MW halo stars would result in surface brightness estimates calculated using the number of stars with $v>v_{\rm M31}$ to be brighter than estimates calculated using the number of stars with $v\le v_{\rm M31}$, and the points would fall below the one-to-one line.  However, the points are consistent with the one-to-one line, indicating that the sample is not significantly contaminated by MW halo stars at faint surface brightnesses (Section~\ref{sec:MWcontamination}). 
}
\label{fig:mu_vel_test}
\end{figure}

Finally, a comparison of the kinematics of the M31 halo sample and
the likely MW halo stars shows that the two have very different distributions at all radii: 
the M31 halo sample remains centered at the systemic velocity of M31, $v_{\rm los}=-300$~\kms,
while the blue MW stars display a broad range of velocities approximately centered at 
the local standard of rest, $v_{\rm lsr}=-175$~\kms.   This is shown quantitatively in Figure~\ref{fig:mu_vel_test}, which displays the surface brightness estimates of a subset of the M31 spectroscopic fields, calculated after splitting the M31 RGB sample at the systemic velocity of M31 ($-300$~\kms).  Incompleteness in the sample, which is expected to occur at velocities overlapping the MW star velocity distribution \citep[Figure~\ref{fig:global_diags},][]{gilbert2007} will cause points to fall above the one-to-one line, as surface brightness estimates calculated using the number of stars with $v>v_{\rm M31}$ will be fainter than estimates calculated using the number of stars with $v\le v_{\rm M31}$.   Conversely, contamination from MW halo stars will cause the points to fall below the one-to-one line.  Since the same velocity PDFs are applied regardless of radius, incompleteness is expected to be roughly uniform, and is implicitly accounted for in the empirical normalization.   Contamination from MW halo stars, however, is likely to be larger at faint surface brightnesses, where the ratio of M31 RGB stars to MW stars is low and the M31 RGB stars have relatively blue colors (Table~\ref{tab:fields}, Figures~\ref{fig:global_diags} and \ref{fig:likelihoods}).  Reassuringly, the data points in Figure~\ref{fig:mu_vel_test} are consistent with the one-to-one line even at faint surface brightnesses.

Combined, these tests indicate that the M31 RGB counts 
are not significantly contaminated by misidentified MW 
stars, and that we have not yet reached the radius at which the surface 
brightness estimates become foreground-limited.

\subsubsection{Beyond the Southern Quadrant}\label{sec:nquad}

Most previous studies of the surface-brightness profile of the outermost 
regions of M31's stellar halo have been confined to the southern quadrant 
or M31's southeast minor axis \citep{guhathakurta2005,irwin2005,ibata2007}.  
\citet{guhathakurta2005} presented a surface brightness profile extending 
out to \rproj\,$\sim 165$~kpc, generated from observations in fields in M31's 
southern quadrant, primarily along M31's SE minor axis.  \citet{irwin2005} 
likewise presented data along M31's SE minor axis to projected distances of \rproj\,$= 55$~kpc.
\citet{ibata2007} presented a 
photometric survey of M31's entire southern quadrant out to \rproj\,$\sim 150$~kpc 
and detected substructure throughout a large portion of this region, much of it related 
to M31's giant southern stream.   These surface brightness profiles were all
consistent with an $r^{-2}$ power-law stellar halo.  

The stars in field m11, located on M31's SE minor axis at a projected radial 
distance of $\sim 165$~kpc from M31's center, are at a projected radial distance 
of $\sim 50$~kpc from the center of M33.  
Although it is more plausible that the stars in field m11 gravitationally belong to M31 
given the relative sizes of the two galaxies, their velocity distribution does 
not rule out the possibility that they belong to M33 \citep{guhathakurta2005}.  
In addition, a large H {\smcap I} bridge between M31 and M33 has been 
observed roughly along M31's minor axis \citep{braun2004},
and a possible interaction between M31 and M33 has been posited based on 
relatively high star count densities observed  extending from M33 in the direction of M31
 \citep[]{mcconnachie2009}.

It could therefore be argued that some or all of the outermost halo M31 
RGB stars (\rproj\,$\gtrsim 100$~kpc) in fields in the southern quadrant 
may be part of large extended substructures connected either with the 
M31/M33 H {\smcap I} bridge or the giant southern stream, and not part 
of a global M31 halo extending to at least $170$~kpc.  If this were the 
case, significant enhancements would be expected in the surface brightnesses
of fields on the SE minor axis and in the S quadrant compared to fields
in one of the other quadrants of M31's stellar halo. The data do not show 
any evidence of such an overdensity (Figure~\ref{fig:sb}).  
 
PAndAS \citep{mcconnachie2009} has used the CFHT
and the MegaCam instrument to survey 
the eastern, northern, and western quadrants of M31's halo
out to a maximum distance of $\sim 150$~kpc. \citet{mcconnachie2009} presented 
the first evidence that the southern quadrant of
M31's stellar halo, although containing large amounts of substructure, still typified
the extent and global structure of the full halo.   Although they did not measure a 
surface brightness profile, they presented a stellar density map showing stars 
consistent with being M31 RGB stars from the SE minor axis 
counter-clockwise to the NW minor axis. They also found significant photometric 
substructure over the full survey area.  \citet{tanaka2010} presented the first 
surface brightness profile composed of stars beyond the southern quadrant, using 
Subaru/Suprime-Cam photometry of contiguous fields along the NW and SE minor axes out to 
\rproj\,$\sim 100$~kpc, and found reasonable consistency between the two profiles (Table~\ref{tab:litfits}). 

The profile presented in Figure~\ref{fig:sb} confirms that M31's stellar halo
follows a consistent power-law profile regardless of position angle.
Eleven of the M31 halo fields are on the northern 
side of the galaxy, spanning a large range of position angle and distance from M31's center 
(Figure~\ref{fig:roadmap}). The surface brightness estimates from these fields
are shown in red in Figure~\ref{fig:sb} and are fully consistent 
with estimates from fields along the SE minor axis and in the southern quadrant.  
Furthermore, secure M31 RGB stars are found in both northern \rproj\,$\sim 170$~kpc 
fields.  This supports the interpretation that the dynamically hot component in the southern quadrant 
fields are part of a global halo population that extends to at least 
170~kpc from M31's center.  

\subsection{Scatter of the Data about the Profile Fit: Effect of Substructure}\label{sec:scatterwithsubstructure}

\begin{figure}[tb!]
\plotone{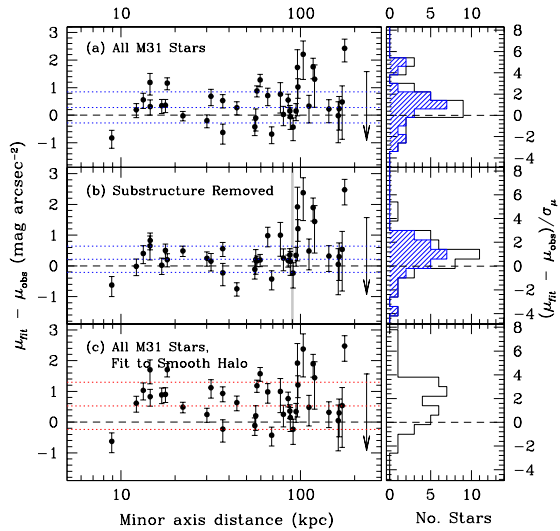}
\caption{Comparison of the scatter of the surface brightness estimates 
about the profiles shown in Figure~\ref{fig:sb}, as a function of distance
from the center of M31 (left panels) and in histogram form (right panels).  
(a) Values of $\Delta\mu$ (defined as $\mu_{\rm fit}-\mu_{\rm obs}$) 
for M31's surface brightness profile including all M31 RGB stars 
(Figure~\ref{fig:sb}(a)).   The histograms show the scatter about the 
best-fit after controlling for the observational errors ($\Delta\mu$/$\sigma_\mu$).
(b) Values of $\Delta\mu$ for the dynamically hot M31 RGB component 
and associated profile fit (Figure~\ref{fig:sb}({\it b})); dynamically 
cold tidal debris features have been statistically removed.  
(c) The aggregate effect of substructure on the profile: 
values of $\Delta\mu$ calculated with the surface brightness 
estimates including all M31 RGB stars (Figure~\ref{fig:sb}(a)) and 
the profile fit to the dynamically hot M31 RGB population 
(Figure~\ref{fig:sb}(b)).  
Shaded histograms are for fields within \rproj\,$<90$~kpc, the outer
limit at which we are able to identify substructure.  In the left panels, the dotted lines 
show the mean and root-mean-square deviation from 
the mean of the $\Delta\mu$ values (blue: calculated for \rproj\,$<90$~kpc, 
red: calculated using all surface brightness estimates).  
Statistical removal of stars associated with dynamically cold tidal debris 
features in M31's stellar halo results in a factor of 1.3 decrease in 
the rms deviation of the data from the best-fit profile (panels a and b, Section~\ref{sec:scatterwithsubstructure}). 
}
\label{fig:delta_mu}
\end{figure}

Although the best-fit power-law indices are consistent for both
surface-brightness profiles shown in Figure~\ref{fig:sb},
the surface brightness estimates including all M31 halo stars show 
slightly more scatter at any given radius (\rproj$\lesssim 90$~kpc) than the estimates that 
include only the fraction of stars belonging to the dynamically hot 
component of M31's stellar halo.  This
is quantified in Figures~\ref{fig:delta_mu}(a) and (b).  Statistical removal of stars
associated with kinematically cold tidal debris features results in a
decrease of a factor of 1.3 in the root-mean-square deviation of the data 
points about the best-fit profile.  The rms value is computed using only data 
at \rproj\,$<90$~kpc since that is the radial limit at which 
kinematical substructure can be detected in the current data set.

The mean enhancement of the surface brightness in a field due to 
kinematically identified substructure is 1 mag; the root-mean-square 
deviation of 
the surface brightness enhancement in fields with kinematically identified 
substructure is 0.5 magnitudes. 
Figure~\ref{fig:delta_mu}(c) shows the aggregate effect of the
inclusion of tidal debris features on the profile; it compares the surface 
brightness estimates including all stars with the best-fit profile to 
the surface brightness estimates with tidal debris statistically subtracted. 

Next, we investigate whether the scatter of the surface 
brightness of individual sight lines in M31's stellar halo increases 
with distance from M31.  Physically, an increase in scatter with 
increasing \rproj\  could be caused by an increasing fraction of stars in 
the outer halo belonging to substructure.  Observationally, an 
increase in scatter with \rproj\ could be due to the smaller number of 
stars observed in the outer fields:  this simultaneously increases the error in the 
surface brightness measurements and reduces our ability to identify and remove
substructure.
  
We can control for increasing error in the surface brightness estimates 
with increasing \rproj\ by analyzing the distribution of $\Delta\mu$/$\sigma_\mu$
values.  If the scatter about the best fit profile is due solely to the observational 
errors, this distribution should have a standard deviation of order unity.  
The standard deviation is computed in three large radial bins: 
\rproj\,$\le30$~kpc, $30<$\,\rproj\,$\le90$~kpc, and \rproj\,$>90$~kpc.  
The standard deviation is $\sim2$ for the fit to fields including substructure, 
regardless of radius.  For the surface brightness
estimates with tidal debris features removed, the standard deviation is $1.5$ in 
both the \rproj\,$\le30$~kpc and $30<$\,\rproj\,$\le90$~kpc bins, and $1.7$ for
fields at \rproj\,$>90$~kpc, where the number of stars per field is insufficient 
to identify kinematically cold components.   

There are several sources of observational error not captured in the formal error 
estimates that could increase the scatter in the surface brightness estimates.
The fraction of stars in tidal debris features is determined by fitting multiple 
Gaussian components to a field's velocity distributions (Section~\ref{sec:sb_est_kcc}).   
However, Gaussians are likely an imperfect approximation of the true velocity 
distribution of the tidal debris: if there are more stars in the tails of the 
velocity distribution of tidal debris features than in the modeled Gaussians, 
this could lead to residual scatter about the profile.  In addition,  
substructure in the MW's halo or a large scale error in the accuracy of the Besan\c{c}on model, 
which is used as a normalization factor, could also increase the scatter in the 
surface brightness estimates (Section~\ref{sec:othererrors}).

\subsection{Shape of M31's Halo}\label{sec:shape}

\begin{figure}[tb!]
\plotone{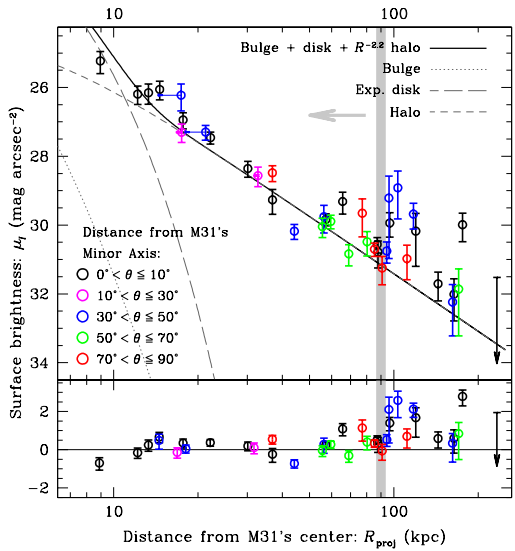}
\caption{Surface brightness profile of M31's stellar halo; data points are color-coded 
by the field's position angle 
from M31's minor axis.
Kinematically identified tidal debris features have been statistically subtracted 
from the surface brightness estimates in fields at \rproj$<90$~kpc.
Open points use \rproj\ for the abscissa, while the abscissa value of the 
smaller filled points (connected by arrows) is their projection onto 
the minor axis assuming a 
flattening of $b/a=0.6$ in M31's inner spheroid 
\citep[\rproj\,$< 30$~kpc; Section~\ref{sec:sbprofile};][]{walterbos1988,pritchet1994}.
There is no apparent systematic offset
of the surface brightness estimates as a function of angular position, indicating that 
the data are consistent with circular isophotes for M31's outer stellar halo (Section~\ref{sec:shape}).  
}
\label{fig:haloflattening}
\end{figure}

The surface brightness estimates cover a large range of projected distances and 
position angles in M31's stellar halo.  
If M31's stellar halo has flattened isophotes with the same orientation as M31's disk, 
data points at increasing position
angles from the minor axis would have systematically higher surface brightness estimates
than points at similar radii but near M31's minor axis.  
If M31 had flattened isophotes with an orientation perpendicular to that of M31's disk,
the data would show systematic offsets in the opposite sense: data points with large
position angles from the disk minor axis would have systematically fainter surface brightness estimates
than data points at similar radii and near M31's minor axis.

We investigate the data for evidence of such an effect in Figure~\ref{fig:haloflattening}.
The surface brightness estimates and best-fit profile are for the sample with
tidal debris features statistically removed to prevent bias of the measurement by strong, large-scale substructure (such as the giant southern stream).
The surface brightness estimates are color-coded by their position angle
from the minor axis (defined such that $\theta\le 90$\degree).  There is no significant, 
systematic offset of any subset of the data 
points from the profile fit, nor does there appear to be an increasing systematic offset with increasing
position angle from the minor axis.  The data appear to be consistent with a
spherical halo in M31.   

We quantify the degree of ellipticity of M31's stellar halo via a maximum-likelihood analysis, under the
constraints that the shape of the halo is assumed to be constant with radius and that the major axis
of M31's halo is aligned with either the major or the minor axis of M31's disk. 
Each data point was projected along M31's (disk) minor axis, 
assuming an axis ratio for the halo isophotes.  The halo axis
ratio is defined with respect to the M31 disk's minor and major axes: a halo axis ratio
smaller than one indicates the halo's major axis is aligned with the disk major axis, while a 
halo axis ratio greater than one indicates the halo's major axis is aligned with the disk minor axis.
The best-fit profile was calculated as described in Section~\ref{sec:fits}.  The maximum-likelihood
halo axis ratio was determined by comparing the $\chi^2$ values of each fit.  
Since the analysis is performed on the sample with kinematically identified substructure 
statistically removed, 
only fields with initial values of \rproj$<90$~kpc are included in the profile fits and
$\chi^2$ measurement; beyond this radius individual fields have too few stars to identify 
kinematical substructure.  

The maximum-likelihood axis ratio for M31's halo is $1.06^{+0.11}_{-0.9}$.  This corresponds to
elliptical isophotes with $b/a=0.94$, with the major axis of the halo aligned along the minor axis of M31's disk, consistent
with a prolate halo.  Circular isophotes (consistent with a spherical halo) are within the 90\% confidence limits, 
which encompass axis ratios of 0.96 to 1.17.  The 99\% confidence limits span a range of halo axis ratios 
of 0.92 and 1.25.  The best-fit power-law index for a halo axis ratio of 1.06 is $-2.2$, 
the same as the power-law index of the nominal fit (Figure~\ref{fig:sb}, Section~\ref{sec:fits}).

Finally, based on previous work \citep{walterbos1988,pritchet1994}, 
we assumed an ellipticity in the surface 
brightness isophotes for fields with \rproj$<30$~kpc for the nominal fits (Section~\ref{sec:sbprofile}).  
In practice, this significantly affected only two of the data points.  
The reader can see the effect of this assumption in 
Figure~\ref{fig:haloflattening}: the open points show the surface 
brightness estimates using the field's \rproj, while the smaller filled points 
(connected with arrows to the \rproj\ points)
show the same surface brightness estimates, but using the effective 
radius along the minor axis assuming an isophotal flattening of $b/a=0.6$ (with the
major axis aligned with the major axis of M31's disk).   
Based on the results of the above analysis, we also tested the profile fits 
using \rproj\ for all data points, rather than assuming flattened isophotes 
in the inner regions, and found fully 
consistent fits with those quoted in Section~\ref{sec:fits}.   
It is possible that the inner regions of M31's stellar halo have a
different ellipticity than the outer regions; however, 
given the limited number of data points at small radii, most of which are near the minor axis, 
we are unable to constrain the shape of M31's inner halo separately
from the shape of the outer halo.

\subsection{Comparison with Previously Published Profiles}\label{sec:compare}

In the last several years multiple measurements of the surface 
brightness profile of M31's outer halo have been made, primarily 
utilizing photometric observations along the southeast minor axis.  
These are summarized in Table~\ref{tab:litfits}.  Although these studies 
largely use independent data sets and analysis, there is partial overlap
in the footprints of several of the surveys; for example, 
seven of the 39 spectroscopic fields presented
here overlap with the \citet{tanaka2010} footprint.   The \citet{courteau2011} analysis
makes use of the \citet{irwin2005} data set and an updated \citet{guhathakurta2005} data set.

\begin{figure*}[tb!]
\centerline{
\includegraphics[width=0.5\textwidth]{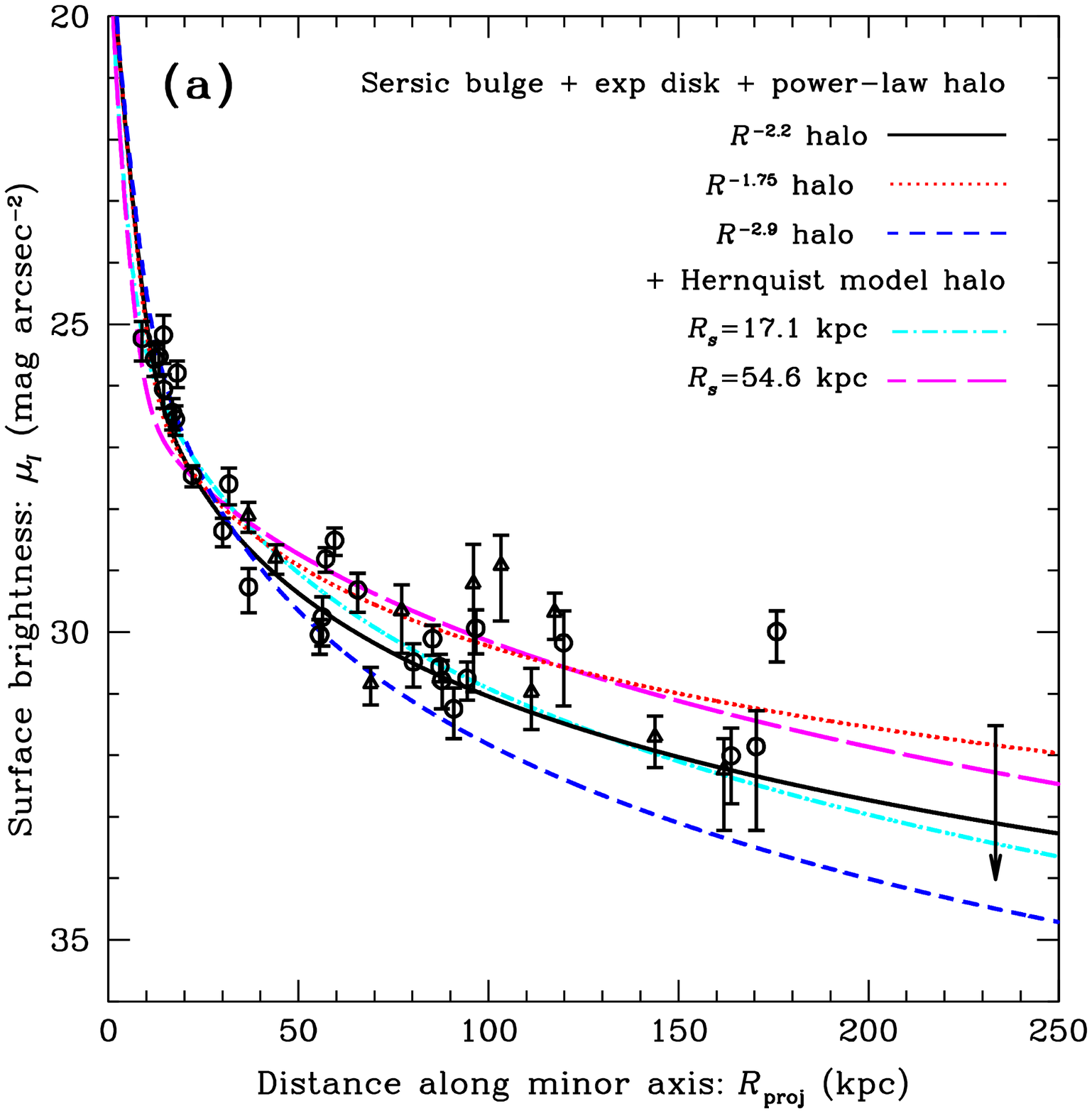}
\includegraphics[width=0.5\textwidth]{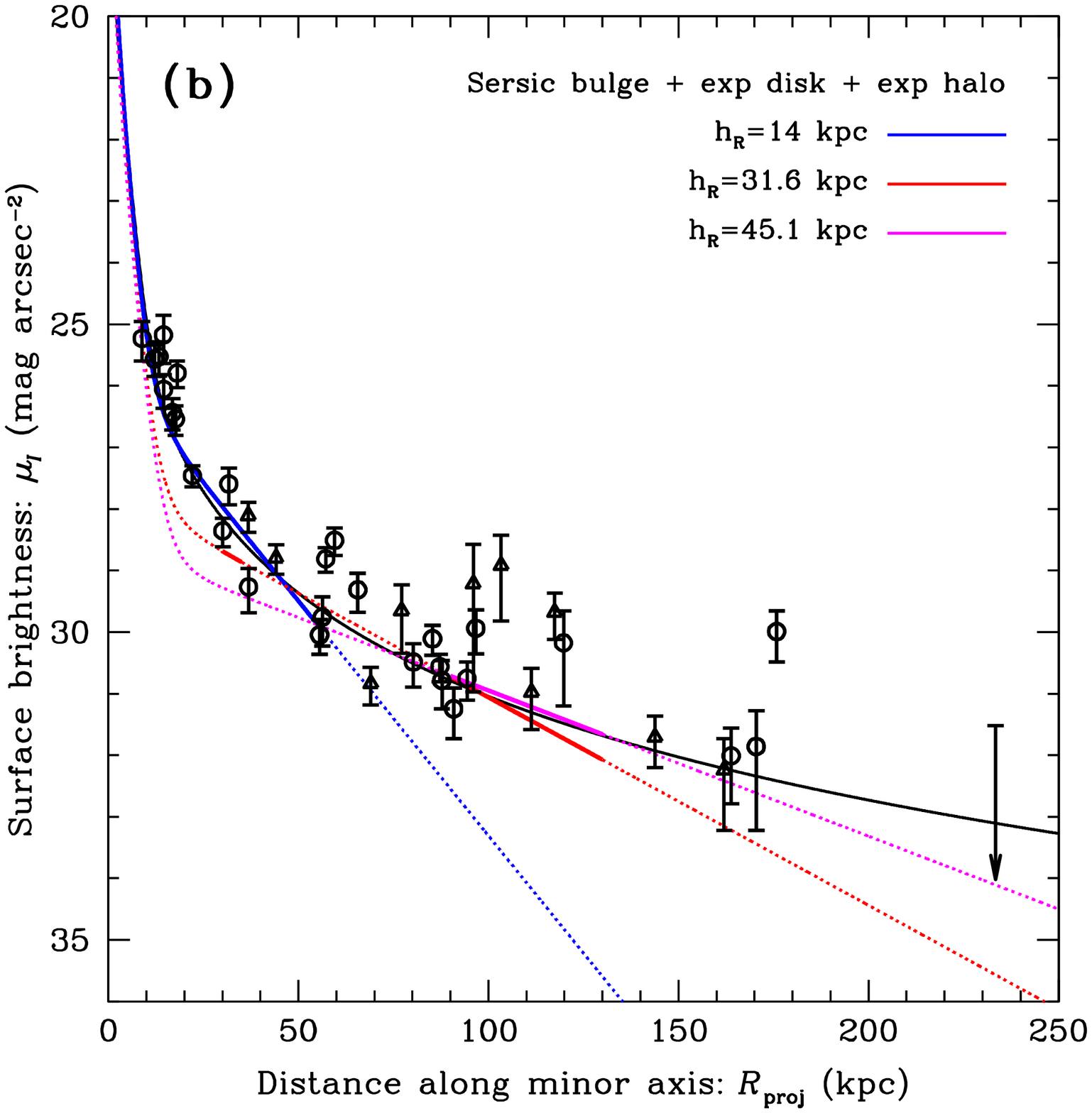}
}
\caption{
Comparison of the surface brightness estimates of 
M31's stellar halo (as in Figure~\ref{fig:sb}(a)) with a range of profile
fits from the literature, in log-linear form.  In
addition to the best-fit power-law profile to the data (black solid curve; 
same as in Figure~\ref{fig:sb}(a)), 
alternative power-law indices and Hernquist models (left) and exponential fits (right) 
have been overlaid (Table~\ref{tab:litfits}, Section~\ref{sec:compare}).  
(a) Alternative power-law indices for the 
outer halo component are overlaid on the data: $-2.9$ (red dotted curve)
and $-1.75$ (blue dashed curve).   Also shown are two Hernquist model fits from
the literature.     
(b) Exponential fits with scale lengths of 14~kpc 
\citep[blue curve;][]{irwin2005}, 
31.6~kpc \citep[red curve;][]{ibata2007}, 
and 45.1~kpc \citep[magenta curve;][]{ibata2007}.  
The exponential fits are shown as solid lines over the radial range in which 
they were measured, and as dotted lines outside this radial range.  Hernquist models are 
consistent with the full range of the data, while exponential profiles are only consistent with the 
data in limited radial ranges. 
}
\label{fig:sb_linear}
\end{figure*}

Figure~\ref{fig:sb_linear} shows the surface brightness estimates 
in log-linear form along with a range of literature fits to 
observations of M31's outer halo 
(\rproj\,$\gtrsim 30$~kpc) surface brightness profile, including 
power-law profiles, Hernquist models (left panel), and exponential profiles (right panel).  
The Hernquist models provide reasonable fits to our data set; the smallest
Hernquist model scale radius from the literature \citep[$R_s=17.1$~kpc,][]{tanaka2010}  
is the most consistent with our data set over its full radial range.
The data are consistent with the exponential fits only within relatively small radial ranges. 
A power-law profile or Hernquist model clearly provides a 
better fit than an exponential profile over the full radial range of the data.

The surface brightness estimates presented here extend to larger projected distances and 
cover a larger azimuthal range in M31's stellar halo than any of the previously 
published profiles.  Our technique of using counts of spectroscopically confirmed 
stars to estimate a surface brightness has an advantage at large projected distances 
over estimates based on photometric observations.  In the sparsest outer regions of 
M31's stellar halo, contaminants from the foreground (MW dwarf stars) and the background 
(distant unresolved galaxies) greatly  
outnumber the target M31 halo population.  Given the large area of sky covered by 
M31's stellar halo (over 20\degree\ in diameter), 
and the strongly varying MW stellar density over this region of sky, the standard technique 
of identifying a comparison field and statistically subtracting it from the science 
image to remove contamination does not work.  Spectroscopic observations 
identify individual objects as foreground MW stars, background galaxies, 
or M31 stars, enabling us to probe M31's halo to large projected distances.

\section{Discussion}\label{sec:discussion}
The density profiles found in simulations of stellar halo formation 
via accretion \citep[e.g.,][]{bullock2005,cooper2010,font2011} are in good
agreement with the measured surface brightness profile of M31's stellar halo.  
\citet{font2011} presented a study of $\approx 400$ stellar halos, simulated using 
a cosmological smoothed particle hydrodynamics code.  Since stellar halo
formation is a somewhat stochastic process, the exact properties of individual
halos do differ.  However,  the median spherically averaged density profiles at large
radii in the \citet{font2011} simulations follow a power-law profile 
with $\rho (r)\propto r^{-3.5}$ 
(leading to a surface density distribution of $\Sigma(r)\propto r^{-2.5}$), in 
good agreement with that measured for the M31 halo surface brightness 
profile.  The \citet{bullock2005} simulations predicted stellar density profiles
that closely followed Hernquist models with scale radii in the range $\sim 10-20$~kpc, also 
in good agreement with the observed surface brightness profile of 
M31 (Figure~\ref{fig:sb_linear}({\it a})).
The published profiles typically
show the aggregate density of the full stellar populations of the simulated 
stellar halos, including tidal streams, and are thus most comparable 
to the observed profile including all M31 stars.  

The data show no evidence for a downward break in the surface brightness profile at large radii.  Most $\Lambda$CDM-based predictions for stellar halo profiles \citep{bullock2001,bullock2005,cooper2010}
suggest that radially-averaged profiles of accreted stellar halos around M31-size galaxies should start to break rapidly around $r\sim 100$~kpc, taking on a steeper profile than that interior to $r\sim100$~kpc.   Of the 11 simulated stellar halos in \citet{bullock2005},
only one, halo 9, did not break to a steeper density profile at large radius; it followed a Hernquist profile with scale radius of 18~kpc, which is similar to the Hernquist scale radius for M31 measured by \citet{tanaka2010} and shown to be consistent with the data in Figure~\ref{fig:sb_linear}.   This simulated halo had an unusually large fraction of recent low-mass accretion/disruption events.  
The large spread in surface brightness in the fields at large radii may indicate that M31 has undergone a similar recent accretion history.
 
Simulations of stellar halo formation via accretion in a 
cosmological context also show that present-day stellar halos should 
contain many photometrically identifiable tidal streams 
\citep[e.g., ][]{bullock2005,johnston2008,cooper2010}. 
Thus, a study utilizing data from isolated sight lines through a stellar halo, 
like this one, should see significant scatter in the surface brightness 
profile due to the presence or absence of tidal streams along each 
line-of-sight.   Furthermore, the simulations find that the contrast 
of stars in coherent tidal debris streams to stars in the diffuse 
stellar halo should become greater at large radii,
as these streams are typically due to more recent accretion events.  
The inner regions of the simulated 
stellar halos tend to be built from early, relatively major mergers, and
are more likely to have undergone significant phase-mixing resulting 
in a smoother stellar distribution.  

The observations presented here provide one of the first direct observational 
tests of this aspect of the simulations.  The surface brightness estimates 
show significant field-to-field variation in the stellar content of M31's halo
at all radii, beyond that expected based on
the estimated observational errors.  Tidal 
debris features are identified via stellar kinematics in $\sim 1/2$ of the fields within 
\rproj\,$\lesssim 90$~kpc.  The tidal debris features generally comprise over 
half of the population in a given field, with the full range of measured substructure fractions extending from 29\% in field d9 to 75\% in field H13s. 
\citet{bullock2001} estimated the expected field-to-field variation in star counts
in stellar halos built via the accretion of satellites
to be a factor of $\sim 2$ to 3 at large radii, in rough agreement with the 
observed field-to-field variation in our data set.  

While removing kinematically identified
substructure reduces the root-mean-square deviation of the data points
about the profile fit, the amount of scatter in the data about the profile fit 
remains greater than that expected from the estimated observational errors, 
suggesting that the outer halo is composed of many small accreted objects that 
are unresolved into kinematically cold clumps or photometrically distinct streams
at the limit of the currently available data.
Interestingly, the data do not show evidence of an increasing fraction of coherent tidal
streams with radius, as expected from simulations and discussed above.  
It is worth noting, however, that the inner
regions of M31's halo contain a high degree of substructure from the recent, relatively
massive accretion event which formed the giant southern stream; in this sense, 
M31 may not be typical.
 
A detailed comparison of the 
fraction of substructure with radius, its effect on the observed surface
brightness profile, and an analysis of the field-to-field variation 
with that predicted in simulations should yield interesting constraints on 
the merger history of M31
and the properties of its former satellite populations (e.g., the LF), 
and may help to provide constraints on the assumptions for satellite properties
used in simulations.
This kind of detailed comparison is currently only possible in M31.  M31's
proximity allows us to obtain detailed observations of the entire 
stellar halo, including 
the crucial kinematical data that allows us to separate tidal debris features
from the smoother underlying stellar halo in individual lines-of-sight.

Interestingly, recent simulations employing smoothed particle
hydrodynamics (and thus explicitly simulating the evolution of the
baryons, including star formation) imply that a significant fraction
of the stellar population in the inner regions of stellar halos may
be formed in the host galaxy potential 
\citep{zolotov2009,font2011,mccarthy2012}. 
This could have several effects on the observed stellar distribution, 
including naturally providing an inner spheroid 
(\rproj\,$\lesssim 10$\,--\,30~kpc) 
with different structural and chemical properties than the outer spheroid, 
 as has been observed in M31.  Indeed, \citet{mccarthy2012} showed
 that in their simulations, the in situ stellar component results in a significantly
 flattened, oblate stellar density distribution in the inner regions 
 (semi-minor to semi-major axis ratio of $b/a\sim 0.3$) that slowly
 becomes more spherical with increasing radius, reaching $b/a\sim 1$ 
 at radii of $\sim 100$~kpc.  The inner regions (\rproj\,$\lesssim 30$~kpc) 
 of M31's spheroid have been shown to have flattened surface brightness contours 
 \citep[$b/a\sim0.6$;][]{walterbos1988,pritchet1994}, while the outer regions
 appear consistent with circular surface brightness contours ($b/a\sim1.0$; this work).
 In contrast, the $N$-body plus semi-analytic
 models, in which star particles are `painted' onto the dark matter distribution,
tend to find a prolate density distribution for the inner stellar halo \citep{cooper2010}. 

Detailed comparisons of the observed distribution of stars in M31's 
stellar halo with simulations of halo formation will be crucial for constraining
the fraction of stars in stellar halos formed in situ. 
Spectroscopic studies such as this one 
will have a natural advantage for such comparisons because the 
spatially diffuse, kinematically hot component can be separated from
and compared to the tidal debris in every field.

\section{Conclusions}\label{sec:conclusion}
We have analyzed data in 38 lines of sight in M31's stellar halo, 
including all quadrants and ranging from 9 to 230~kpc in projected
distance from M31's center.  Spectroscopic observations
of stars were obtained in a total of 108 multi-object slit masks.  The stellar
spectra were used to identify samples of M31 RGB stars
and  MW dwarf star contaminants, and to identify M31 stars belonging
to tidal debris features.  Counts of M31 RGB stars and MW dwarf 
stars were used to determine the surface brightness in each line-of-sight, 
yielding a surface brightness profile of M31's stellar halo.

We expand upon our main conclusions below.  In brief, our conclusions are as follows.
\begin{itemize}
\item We present a clear detection of M31's stellar 
halo to projected distances of 175~kpc from M31's center.
\item Multiple tests indicate there is minimal contamination from misclassified foreground MW stars 
in the sample of M31 RGB stars.
\item The surface brightness profile follows a power-law with an index of 
$-2.2$.  
\item The statistical subtraction of tidal debris allows us
to investigate its effect on the measured surface brightness profile.  
\end{itemize}

{\it We present a clear detection of M31's stellar 
halo to projected distances of 175~kpc from M31's center (Figure~\ref{fig:sb}).}  
Our data set includes lines-of-sight spread 
throughout all quadrants of M31's stellar halo, covering a range of radii.  
The data show that M31's stellar halo 
extends to at least $\sim 2/3$ of M31's estimated virial radius 
\citep[$\sim 260$~kpc;][]{seigar2008}.
In aggregate, a large spread in the velocities of M31 halo stars
is observed even in fields at large projected distances from M31's center 
(\rproj$>100$~kpc, Figure~\ref{fig:likelihoods}).    

{\it Multiple diagnostic tests indicate there is minimal contamination from misclassified foreground MW stars 
in the sample of M31 RGB stars.}  
The spectroscopic data enable us to remove foreground and 
background contaminants, allowing us to trace 
M31's stellar halo to fainter surface brightnesses, and larger radii, than 
studies based on photometry alone.
Tests show that using the \citet{gilbert2006} likelihood method to identify M31 
RGB and MW dwarf stars does an excellent job of cleanly separating the two 
populations (Section~\ref{sec:MWcontamination}, Figures~\ref{fig:mu_gallat} and \ref{fig:mu_vel_test}).  
If the M31 RGB sample was significantly contaminated by MW disk stars, 
there would be a strong dependence on the surface brightness of fields 
as a function of Galactic latitude. 
No such dependence is seen in the data.
The density of MW halo stars is expected to be relatively uniform 
across the face of M31, 
so significant contamination would lead to a constant surface brightness 
measurement.  
Our sample instead displays a systematic fall off with projected distance 
from M31's center.   Furthermore, surface brightness estimates in individual 
fields calculated by splitting the M31 sample at the systemic velocity of M31 are
consistent even at faint surface brightnesses, indicating that contamination from MW halo
stars is not significant.  Even at large radii (\rproj$>100$~kpc) there are
stars with properties consistent with M31 RGB stars rather than MW halo or 
disk stars (Figures~\ref{fig:global_diags} and \ref{fig:likelihoods}).

{\it The surface brightness profile follows a power law with an index of 
$-2.2$.}  
This is consistent with what previous studies 
that have mainly targeted the southern quadrant of M31 have found.  This work 
establishes that an $\sim r^{-2}$ power-law profile is a good description of 
M31's outer halo (\rproj\,$\gtrsim$~20\,--\,30~kpc) regardless of quadrant or radius.  
The data show no evidence of a downward break in the profile at large radii (\rproj\,$\gtrsim100$~kpc).
Since the fields cover a large range in both radii and position angle around M31,
they provide leverage on the shape of M31's halo.  The best-fit elliptical isophotes
have $b/a=0.94$, with the major axis of the halo aligned along the minor axis of M31's disk, 
consistent with a prolate halo.  However, a value of $b/a=1$ is well within the 90\% confidence limits 
of the fit, so the data are consistent with a spherical halo.

{\it The statistical subtraction of tidal debris allows us
to investigate its effect on the measured surface brightness profile.}  
The ability to remove stars associated with tidal debris from 
individual lines of sight is a unique advantage of our spectroscopic survey
over photometric surveys of M31's stellar halo.  
There are enough stars per field to discover kinematically cold tidal debris 
features for fields with \rproj\,$\lesssim 90$~kpc.  
Tidal debris features are identified using kinematics in $\sim 1/3$ 
of the spectroscopic fields ($\sim 1/2$ of fields with 
\rproj\,$\lesssim 90$~kpc). 
In fields with substructure, the mean enhancement of the surface brightness 
is 1 mag, with a root-mean-square deviation of the sample about the 
best-fit profile of 0.5 mag. 
The scatter about the mean surface brightness profile decreases 
once kinematically identified tidal debris features are 
statistically subtracted. 
This implies that there is a comparatively diffuse stellar component 
to M31's stellar halo out to at least \rproj\,$\sim 90$~kpc, with relatively prominent 
tidal debris features superimposed.

\acknowledgments
We thank Teresa Krause of Castilleja School, a participant in the Science Internship 
Program (SIP) at UCSC, for her work on the separation of M31 RGB and MW dwarf star samples.  
Support for this work was provided by NASA through Hubble Fellowship 
grants 51273.01 and 51256.01 awarded to K.M.G. and E.N.K. by the Space Telescope 
Science Institute, which is operated by the Association of Universities for 
Research in Astronomy, Inc., for NASA, under contract NAS 5-26555.\\
P.G., J.S.B., and S.R.M. acknowledge support from collaborative NSF grants AST-1010039, AST-1009973, AST-1009882, and AST-0607726.
This project was also supported by NSF grants AST03-07842, AST03-07851, AST06-07726, AST08-07945 and AST10-09882, NASA grant HST-GO-12105.03 through STScI, NASA/JPL contract 1228235, the David and Lucile Packard Foundation, and the F. H. Levinson Fund of the Peninsula Community Foundation (S.R.M., R.J.P., and R.L.B.).
E.J.T. acknowledges support from a Graduate Assistance in Areas of National Need (GAANN) fellowship. R.L.B. acknowledges receipt of the Mark C. Pirrung Family Graduate Fellowship from the Jefferson Scholars Foundation and a Fellowship Enhancement for Outstanding Doctoral Candidates from the Office of the Vice President of Research at the University of Virginia. 
The analysis pipeline used to reduce the
DEIMOS data was developed at UC Berkeley with support from NSF grant
AST-0071048.

The authors recognize and acknowledge the very significant cultural role 
and reverence that the summit of Mauna Kea has always had within the indigenous Hawaiian community. We are most fortunate to have the opportunity to conduct observations from this mountain.


\bibliography{m31}


\begin{deluxetable}{lcccrrrccl}
\tablecolumns{10}
\tablecaption{Basic Results of Spectroscopic Observations.}
\tablehead{\multicolumn{1}{c}{Field} & \multicolumn{1}{c}{No.} & \multicolumn{1}{c}{Mean} & \multicolumn{1}{c}{Filters\tablenotemark{b}} & \multicolumn{1}{c}{No.}  & \multicolumn{1}{c}{No.} & \multicolumn{1}{c}{No.} &  \multicolumn{1}{c}{Kinematically} & \multicolumn{1}{c}{Dwarf} & \multicolumn{1}{c}{Reference\tablenotemark{e}} \\
 & \multicolumn{1}{c}{Masks} & \multicolumn{1}{c}{Projected} &  & \multicolumn{1}{c}{Stellar} &  \multicolumn{1}{c}{Secure} & \multicolumn{1}{c}{Secure} & \multicolumn{1}{c}{Identified} & \multicolumn{1}{c}{Satellite} &  \\
   &   & \multicolumn{1}{c}{Radius} & &  \multicolumn{1}{c}{Spectra\tablenotemark{c}} & \multicolumn{1}{c}{M31} & \multicolumn{1}{c}{MW} & \multicolumn{1}{c}{Substructure}  & \multicolumn{1}{c}{Field} &  \\
   &   & \multicolumn{1}{c}{(kpc)\tablenotemark{a}} & &  & \multicolumn{1}{c}{RGB} & \multicolumn{1}{c}{Dwarf} & & & \\ 
   &   &  & &  & \multicolumn{1}{c}{Stars\tablenotemark{d}} & \multicolumn{1}{c}{Stars\tablenotemark{d}} & & & 
}
\startdata
f109    & 1 &  9  & $g'$, $i'$ & 186 & 148  &  13 &     &    & (1) \\ 
H11     & 2 & 12  & $g'$, $i'$ & 206 & 165  &  21 & Yes &    & (1) \\ 
f116    & 1 & 13  & $g'$, $i'$ & 171 & 140  &  18 & Yes &    & (1) \\ 
f115    & 1 & 15  & $g'$, $i'$ & 136 &  95  &  20 &     &    & (1) \\ 
f207    & 1 & 17  & $g'$, $i'$ &  123 & 113  &   9 & Yes &    & (2) \\ 
f135    & 1 & 17  & $g'$, $i'$ & 133 &  93  &  24 & Yes &    & (1) \\ 
f123    & 1 & 18  & $g'$, $i'$ & 126 &  96  &  28 & Yes &    & (1) \\ 
H13s    & 2 & 21  & $g'$, $i'$ & 212 & 177  &  30 & Yes &    & (3) (2) \\ 
f130    & 3 & 22  & $g'$, $i'$ & 201 &  98  &  70 &     &    &  (1) \\ 
a0      & 3 & 30  & $M$, $T_2$, DDO51 & 120 &  70 &  34 &     &    & (1) \\ 
a3      & 3 & 33  & $M$, $T_2$, DDO51 &  88 &  78 &  17 & Yes &    & (4) (2) \\ 
mask4   & 1 & 37 & $g'$, $i'$ & 57 & 13 & 37 &     &    & \\
d9      & 2 & 37 & $M$, $T_2$, DDO51 & 133 & 46 & 38 & Yes & And~IX & (5) \\ 
d1      & 2 & 44 & $M$, $T_2$, DDO51 & 185 & 44 & 41 & Yes & And~I & (2) (6) \\ 
m4      & 5 & 57  & $M$, $T_2$, DDO51 & 160 & 46 &  93 & Yes &    & (2) \\ 
R04A240 & 3 & 56 & $M$, $T_2$, DDO51 & 123 & 19 & 87 &     &    & \\
R04A338 & 3 & 56 & $M$, $T_2$, DDO51 & 101 &  9 & 72 &     &    & \\
a13     & 4 & 58  & $M$, $T_2$, DDO51 & 114 & 42 &  56 & Yes &    & (2) \\ 
streamE & 2 & 66 & $V$, $I$ & 117 & 15 & 71 &     &    & \\
d3      & 3 & 69 & $M$, $T_2$, DDO51 & 192 & 15 & 104 &     & And~III & (2) (6) (5) \\ 
d10     & 2 & 77 &  $V$, $I$ & 100 & 5 & 49 &     & And~X & (7) (6) (5) \\ 
a19     & 4 & 80  & $M$, $T_2$, DDO51 & 142 &  11 & 115 &     &    &  \\
R06A220 & 3 & 85 & $M$, $T_2$, DDO51 & 146 & 26 & 102 & Yes &    & \\
m6      & 5 & 87  & $M$, $T_2$, DDO51 & 235 &  28  & 174 &     &    &  \\
R06A310 & 3 & 88 & $M$, $T_2$, DDO51 & 169 &  9 & 135 &     &    & \\
R06A040 & 3 & 91 & $M$, $T_2$, DDO51 & 197 & 8 & 161 &     &    & \\
b15     & 5 & 94  & $M$, $T_2$, DDO51 & 144 &  15 & 103 &     &    & \\
d12     & 1 & 96 & $M$, $T_2$, DDO51 & 12 & 2 & 7 &     & And~XII & (5) \\ 
streamF & 3 & 97 & $V$, $I$ & 170 & 11 & 118  &      &    &  \\
d11     & 1 & 103 & $M$, $T_2$, DDO51 & 19 & 4 & 14 &     & And~XI & (5) \\ 
d5      & 3 & 111 & $M$, $T_2$, DDO51 & 167 & 6 & 56 &     & And~V & (5) \\ 
d13     & 5 & 117 & $M$, $T_2$, DDO51 & 62 & 14 & 26 &     & And~XIII & (5) \\ 
m8      & 2 & 120 & $M$, $T_2$, DDO51 &  36 &   3 &  25 &     &    & \\
d2      & 11 & 144 & $M$, $T_2$, DDO51 & 654 & 8 & 98 &     & And~II & (6) (5) \\ 
R11A170 & 3 & 162 & $M$, $T_2$, DDO51 & 117 &   3 &  45 &     & And~XIV & (8) (6) (5) \\ 
m11     & 4 & 164 & $M$, $T_2$, DDO51 & 117 &   4 &  93 &     &    & \\
R11A080    & 2 & 170 & $M$, $T_2$, DDO51 & 127 & 2 & 103 &     &    &  \\
R11A305    & 2 & 176 & $M$, $T_2$, DDO51 & 147 & 8 & 123 &     &    &  \\
d7      & 2 & 233 & $M$, $T_2$, DDO51 & 217 & 0 & 26 &     & And~VII & (6) (5) \\ 
\enddata
\label{tab:fields}
\tablenotetext{a}{The distance of the field from M31's center, obtained by calculating the mean distance from M31 of all stellar spectra in a field.}
\tablenotetext{b}{Bandpasses of the photometric catalog used to design the spectroscopic masks.}
\tablenotetext{c}{Number of unique stellar spectra with successful velocity measurements, excluding duplicate measurements and alignment stars.}
\tablenotetext{d}{These columns report the number of stars designated as M31 RGB halo stars or MW dwarf stars as determined by the diagnostic method described in \S\ref{sec:cleansample}.  In dwarf galaxy fields, RGB stars identified as likely dSph members are not included in the M31 RGB count (Section~\ref{sec:dsph_fields})}
\tablenotetext{e}{Listed references discuss in detail the Keck/DEIMOS spectroscopic observations in the field: (1) \citet{gilbert2007}; (2) \citet{gilbert2009gss}; (3) \citet{kalirai2006gss}; (4) \citet{guhathakurta2006}; (5) \citet{tollerud2012}; (6) \citet{kalirai2010}, (7) \citet{kalirai2009}; (8) \citet{majewski2007} }
\end{deluxetable}

\begin{deluxetable}{lrrrlll}
\tablecaption{Surface Brightness Estimates Based on a Spectroscopically Confirmed Sample.}
\tablehead{\multicolumn{1}{c}{Field} & \multicolumn{1}{c}{R.A.\tablenotemark{a}} & \multicolumn{1}{c}{Decl.\tablenotemark{a}} & \multicolumn{2}{c}{$\mu_I$\tablenotemark{b}} \\
 & \multicolumn{1}{c}{(hh:mm:ss)} & \multicolumn{1}{c}{(\degree:\arcmin:\arcsec)} & \multicolumn{2}{c}{(mag arcsec$^{-2}$)} \\ 
 &   \multicolumn{1}{c}{(J2000)}         &        \multicolumn{1}{c}{(J2000)}                & \multicolumn{1}{c}{All} & \multicolumn{1}{c}{Substructure} \\
 &            &                        & \multicolumn{1}{c}{Stars} & \multicolumn{1}{c}{Removed\tablenotemark{c}} \\}
\startdata
f109 &  00:45:46.75 &  40:56:53.8 & $25.23^{+0.37}_{-0.28}$ & \\ 
H11 & 00:46:21.02  & 40:41:31.3 & $25.56^{+0.29}_{-0.23}$ & $26.19^{+0.30}_{-0.24}$ \\ 
f116 & 00:46:54.53 & 40:41:29.5 & $25.52^{+0.31}_{-0.24}$ & $26.15^{+0.33}_{-0.25}$ \\ 
f115 & 00:47:32.71  & 40:42:00.9 & $26.06^{+0.31}_{-0.24}$ & \\ 
f207 & 00:43:42.64 & 40:00:31.6  & $25.17^{+0.46}_{-0.32}$  & $26.22^{+0.46}_{-0.32}$  \\
f135 & 00:46:24.88  & 40:11:35.5 & $26.44^{+0.28}_{-0.22}$ & $27.30^{+0.30}_{-0.23}$ \\ 
f123 & 00:48:05.57 &  40:27:16.3 & $26.54^{+0.26}_{-0.21}$ & $26.94^{+0.26}_{-0.21}$ \\ 
H13s & 00:44:14.76 &  39:44:18.2 & $25.79^{+0.24}_{-0.20}$ & $27.30^{+0.23}_{-0.19}$ \\ 
f130 & 00:49:08.02 & 40:14:39.8 & $27.46^{+0.18}_{-0.16}$ & \\ 
a0 & 00:51:43.79 & 39:53:47.7 & $28.36^{+0.25}_{-0.21}$ & \\ 
a3 & 00:48:10.52 & 39:07:04.7 & $27.59^{+0.34}_{-0.26}$ & $28.56^{+0.32}_{-0.25}$ \\ 
mask4 & 00:54:08.34  & 39:41:51.7 & $29.27^{+0.42}_{-0.30}$ & \\ 
d9 & 00:52:50.58 & 43:11:09.0 & $28.11^{+0.27}_{-0.22}$ & $28.48^{+0.26}_{-0.21}$ \\ 
d1 & 00:46:01.28 & 38:03:06.9 & $28.80^{+0.27}_{-0.21}$ & $30.18^{+0.24}_{-0.20}$ \\ 
m4 & 01:00:10.49 & 38:49:45.7 & $28.81^{+0.22}_{-0.18}$ & $29.83^{+0.19}_{-0.16}$ \\ 
R04A240 & 00:24:05.34 & 39:21:06.3 & $30.05^{+0.32}_{-0.24}$ & \\ 
R04A338 & 00:34:57.17 & 45:07:26.8 & $29.76^{+0.47}_{-0.33}$ & \\ 
a13 & 00:42:06.13 & 37:01:40.3 & $28.51^{+0.25}_{-0.20}$ & $29.89^{+0.21}_{-0.18}$ \\ 
streamE & 00:22:11.76 & 44:11:57.9 & $29.31^{+0.36}_{-0.27}$ & \\ 
d3 & 00:35:51.72 & 36:24:34.6 & $30.84^{+0.35}_{-0.26}$ & \\ 
d10 & 01:06:37.66 & 44:49:04.6 & $29.65^{+0.69}_{-0.42}$ & \\ 
a19 & 00:37:30.62 & 35:33:28.1 & $30.48^{+0.41}_{-0.30}$ & \\ 
R06A220 & 00:22:23.41 & 36:25:30.5 & $30.11^{+0.27}_{-0.22}$ & $30.70^{+0.21}_{-0.18}$ \\ 
m6 & 01:08:55.43 & 37:35:59.3 & $30.56^{+0.25}_{-0.20}$ & \\ 
R06A310 & 00:16:10.87 & 45:25:59.3 & $30.79^{+0.46}_{-0.32}$ & \\ 
R06A040 & 01:06:22.60 & 46:18:29.6 & $31.25^{+0.49}_{-0.34}$ & \\ 
b15 & 00:53:14.13 & 34:52:38.6 & $30.75^{+0.35}_{-0.27}$ & \\ 
d12 & 00:47:35.72  & 34:20:59.0 & $29.21^{+1.76}_{-0.64}$ & \\ 
streamF & 00:12:05.81 & 45:33:48.1 & $29.94^{+0.41}_{-0.30}$ & \\ 
d11 & 00:46:29.15 &  33:50:05.0 & $28.91^{+0.91}_{-0.49}$ & \\ 
d5 & 01:10:21.00 & 47:39:11.7 & $30.98^{+0.61}_{-0.39}$ & \\ 
d13 & 00:51:56.75 & 33:00:22.0 & $29.68^{+0.44}_{-0.31}$ & \\ 
m8 & 01:18:23.72 & 36:15:27.9 & $30.18^{+1.03}_{-0.52}$ & \\ 
d2 & 01:16:25.19 & 33:26:10.7 & $31.71^{+0.50}_{-0.34}$ & \\ 
A170 & 00:51:41.82 & 29:40:49.4 & $32.24^{+0.98}_{-0.51}$ & \\ 
m11 & 01:29:56.91 & 34:17:10.9 & $32.01^{+0.78}_{-0.45}$ & \\ 
R11A080 & 01:47:09.12 & 43:20:24.0 & $31.86^{+1.36}_{-0.59}$ & \\ 
R11A305 & 23:44:10.24 & 48:05:56.0 & $29.99^{+0.49}_{-0.34}$ & \\ 
\enddata
\label{tab:sbest}
\tablenotetext{a}{For fields in which only one mask was observed or all mask centers overlapped, the R.A. and decl. given is the center of the spectroscopic mask(s).  For the rest of the fields, the R.A. and decl. are the mean R.A. and decl. of all observed masks.}  
\tablenotetext{b}{The normalization of the surface brightness estimates is based on \citet{courteau2011} (Section~\ref{sec:sbprofile}).}
\tablenotetext{c}{Empty values indicate that no kinematic substructure was detected in the field.}
\end{deluxetable}

\begin{deluxetable}{lrrrlll}
\tablecolumns{6}
\tablewidth{0pc}
\tablecaption{Recent Profile Measurements of M31's Halo.}
\tablehead{\multicolumn{1}{c}{Study} & \multicolumn{1}{c}{$R_{\rm min}$} & \multicolumn{1}{c}{$R_{\rm max}$} & \multicolumn{1}{c}{Power-law} & \multicolumn{1}{c}{Data type} & \multicolumn{1}{c}{Notes}  \\
 & \multicolumn{1}{c}{(kpc)} & \multicolumn{1}{c}{(kpc)} &  \multicolumn{1}{c}{Index} &  \multicolumn{1}{c}{(Fit Type)} & \\
 &  &  &  \multicolumn{1}{c}{(Scale Length)\tablenotemark{a}} &  & \\
}
\startdata
\citet{guhathakurta2005}    & 30 &  165  & $-2.3$ & Spectroscopic & S quadrant \\
                            &     &       &        &      & (included 10 of the fields presented here) \\
\citet{irwin2005}\tablenotemark{b}           & 20 &  55   & $-2.3$ & Photometric   & SE minor axis \\
                            &     &       &  (14)  &      (Exponential) & \\
\citet{ibata2007}\tablenotemark{c}           & 30-35 &  90-130\tablenotemark{d}  & $-1.9\pm0.12$& Photometric   & SE minor axis \\
                            &     &       &  ($31.6\pm1.0$)  &      (Exponential) &  \\
                            &     &       &  ($54.6\pm1.3$)  &      (Hernquist) & \\
                            &  90 &  130  &  ($45.1\pm6.0$)  &      (Exponential) & \\
                            &     &       &  ($53.1\pm3.5$)  &       (Hernquist) & \\
\citet{tanaka2010}\tablenotemark{c,e}          & 20 &  100 & $-1.75\pm0.13$ & Photometric   & SE minor axis \\
                            &     &       &  ($22.4\pm2.3$)  &  (Exponential)    &  \\
                            &     &       &  ($31.7\pm6.7$)  &    (Hernquist)  &  \\
                            & 20 &  100  & $-2.17\pm0.15$ &      & NW minor axis \\
                            &     &       &  ($18.8\pm1.8$)  &   (Exponential)   &  \\
                            &     &       &  ($17.1\pm4.7$)  &  (Hernquist)    &  \\
\citet{courteau2011}\tablenotemark{f}        & 0 &  165  & $-2.5\pm0.2$ & Photometric and  & SE minor axis preferred fit \\
                            &     &       &     & Spectroscopic & (included 11 of the fields presented here) \\
\citet{williams2012}  & 2  & 35 & $-2.6^{+0.3}_{-0.2}$ & Photometric & BHB stars used as tracers of the halo \\ 
\\
\hline
\\
This work          & 9 &  176  & $-2.2\pm0.2$ & Spectroscopic & All quadrants, all M31 stars \\
                           & 35 &  176  & $-1.9\pm0.4$ &      & All quadrants, all M31 stars \\
                            & 9 &   90  & $-2.2\pm0.3$ &     & Tidal debris statistically subtracted \\

\enddata
\label{tab:litfits}
\tablenotetext{a}{The scale radii (in kpc) of exponential fits and Hernquist models are denoted within parentheses.}
\tablenotetext{b}{The \citet{irwin2005} study contained integrated light measurements and RGB star count measurements of M31's surface brightness from M31's center out to 55~kpc.  Only the data from 20~kpc outward was used to determine the profile of the outer halo.}
\tablenotetext{c}{The profile fit is to metal-poor (\feh\,$<-0.7$) stars only. }
\tablenotetext{d}{The profile is fit to stars in two radial regions.  The data between 35 and 90 kpc were excluded from this analysis due to numerous tidal debris features.}
\tablenotetext{e}{Similar to \citet{ibata2007}, regions deemed overdense (i.e., strongly contaminated by tidal debris features) were excluded from the profile fit. }
\tablenotetext{f}{The \citet{courteau2011} study combined integrated light measurements \citep{choi2002}, photometry-based RGB star counts (from \citet{pritchet1994} and \citet{irwin2005}), and spectroscopy-based RGB star counts (11 of the fields presented here).  They simultaneously fit for the bulge, disk, and outer halo profiles.  Five of their data points were at \rproj\,$>55$~kpc. }
\end{deluxetable}

\end{document}